\documentclass[a4paper,UKenglish,cleveref, autoref, thm-restate]{lipics-v2021}
\hideLIPIcs  %uncomment to remove references to LIPIcs series (logo, DOI, ...), e.g. when preparing a pre-final version to be uploaded to arXiv or another public repository

\title{Minimum Membership Dominating Set and Its 
Generalization: Leafage vs.\ Asteroidal Number} %TODO Please add

\titlerunning{Minimum Membership Dominating Set and Its 
Generalization} %TODO optional, please use if title is longer than one line

\author{Yuma Tamura}{Graduate School of Information Sciences, Tohoku University, Sendai, Japan}{tamura@tohoku.ac.jp}{https://orcid.org/0009-0001-5479-7006}{JSPS KAKENHI Grant Numbers JP24H00686 and JP25K21148.}

\author{Daichi Wakayama}{Graduate School of Information Sciences, Tohoku University, Sendai, Japan}{daichi.wakayama.p8@dc.tohoku.ac.jp}{}{}

\author{Akira Suzuki}{Center for Data-Driven Science and Artificial Intelligence, Tohoku University, Sendai, Japan \and \url{https://www.ecei.tohoku.ac.jp/alg/suzuki/}}{akira@tohoku.ac.jp}{https://orcid.org/0000-0002-5212-0202}{JSPS KAKENHI Grant Number JP25K14980.}

\author{Xiao Zhou}{Graduate School of Information Sciences, Tohoku University, Sendai, Japan}{zhou@tohoku.ac.jp}{}{}

\authorrunning{Y. Tamura, D. Wakayama, A. Suzuki, and X. Zhou} %TODO mandatory. First: Use abbreviated first/middle names. Second (only in severe cases): Use first author plus 'et al.'

\Copyright{Yuma Tamura, Daichi Wakayama, Akira Suzuki, and Xiao Zhou} %TODO mandatory, please use full first names. LIPIcs license is "CC-BY";  http://creativecommons.org/licenses/by/3.0/

 %TODO mandatory; please add comma-separated list of keywords

\relatedversion{} %optional, e.g. full version hosted on arXiv, HAL, or other respository/website
\acknowledgements{We would like to thank the anonymous referees for their helpful comments.}%optional

\nolinenumbers %uncomment to disable line numbering

\EventEditors{Tuukka Korhonen and Michael Lampis}
\EventNoEds{2}
\EventLongTitle{21st International Symposium on Parameterized and Exact Computation (IPEC 2026)}
\EventShortTitle{IPEC 2026}
\EventAcronym{IPEC}
\EventYear{2026}
\EventDate{September 2--4, 2026}
\EventLocation{L’Aquila, Italy}
\EventLogo{}
\SeriesVolume{398}
\ArticleNo{25}
\usepackage{graphicx, xcolor}
\usepackage{mathrsfs}
\usepackage{ascmac}
\usepackage{tikz}
\usetikzlibrary{calc,positioning,shapes}
\usetikzlibrary{decorations.pathreplacing}

\definecolor{figgray}{HTML}{F0F0F0}
\definecolor{figred}{HTML}{DA6A76}
\definecolor{figgreen}{HTML}{90C482}
\definecolor{figblue}{HTML}{4CA6DE}
\definecolor{figedge}{HTML}{00B050}
\definecolor{figvertex}{HTML}{0070C0}
\definecolor{figheight}{HTML}{FF0000}

\tikzset{
  figure/.style={x=.5pt, y=-.5pt, line width=.7pt, line join=round},
  gadget figure/.style={figure, line width=1.2pt},
  figlabel/.style={scale=#1, inner sep=0pt},
  figlabel/.default=2.6,
  model/.style={line width=4pt, line cap=butt, line join=miter, color=#1},
}

\newcommand{\figcircle}[3]{\draw[fill=figred] (#1,#2) circle[radius=#3];}
\newcommand{\figtriangle}[3]{\draw[fill=figgreen] (#1,#2) ++(-1.2*#3,.95*#3) -- ++(2.4*#3,0) -- ++(-1.2*#3,-2.05*#3) -- cycle;}
\newcommand{\figsquare}[3]{\draw[fill=figblue] (#1-#3,#2-#3) rectangle (#1+#3,#2+#3);}
\newcommand{\figdiamond}[3]{\draw[fill=white] (#1,#2-#3) -- (#1+#3,#2) -- (#1,#2+#3) -- (#1-#3,#2) -- cycle;}
\newcommand{\figstar}[3]{\node[draw, fill=white, star, star points=5, star point ratio=2.3, inner sep=0pt, minimum size=#3*1pt] at (#1,#2) {};}

\newcommand{\figterminal}[4]{%
  \draw[fill=white] (#1-#3,#2-#4) rectangle (#1+#3,#2+#4);
  \draw (#1-#3,#2-.78*#4) -- (#1+#3,#2-.78*#4);
  \draw (#1-#3,#2+.78*#4) -- (#1+#3,#2+.78*#4);
}

\newcommand{\figchain}[5]{%
  \draw (#1,#2) -- (#1,#2+9*#3);
  \foreach \figi in {0,1,2}{
    \pgfmathsetmacro{\figa}{#2+3*\figi*#3}
    \pgfmathsetmacro{\figb}{\figa+#3}
    \pgfmathsetmacro{\figc}{\figb+#3}
    \if e#5
      \figcircle{#1}{\figa}{#4}
      \figtriangle{#1}{\figb}{#4}
    \else
      \figtriangle{#1}{\figa}{#4}
      \figcircle{#1}{\figb}{#4}
    \fi
    \figsquare{#1}{\figc}{#4}
  }
  \figterminal{#1}{#2+9.24*#3}{1.12*#4}{1.68*#4}
}

\newcommand{\figselection}[4]{%
  \pgfmathsetmacro{\figw}{(#3)-(#1)}
  \pgfmathsetmacro{\figh}{(#4)-(#2)}
  \pgfmathsetmacro{\figm}{(#2+#4)/2}
  \draw[fill=white, rounded corners=.1*\figh*1pt] (#1,#2) rectangle (#3,#4);
  \foreach \figp in {.16,.39,.85}{
    \draw (#1+\figp*\figw,\figm) circle[radius=.29*\figh];
  }
  \fill (#1+.58*\figw,\figm) circle[radius=.03*\figh] (#1+.62*\figw,\figm) circle[radius=.03*\figh] (#1+.66*\figw,\figm) circle[radius=.03*\figh];
}
\newcommand{\figbound}[6]{%
  \draw[fill=white, rounded corners=6pt] (#1,#2) rectangle (#3,#4);
  \draw[rounded corners=4pt] (#1+#5,#2+#5) rectangle (#3-#5,#4-#5);
  \node[figlabel=#6] at ({(#1+#3)/2},{(#2+#4)/2}) {\figboundlabel};
}
\newcommand{\figattached}[2]{\draw[fill=white] (#1-30,#2-30) rectangle (#1+30,#2+30);}
\newcommand{\figarrow}[2]{\draw[fill=white] (#1,#2-22) -- ++(66,0) -- ++(0,-22) -- ++(44,44) -- ++(-44,44) -- ++(0,-22) -- ++(-66,0) -- cycle;}

\newcommand{\figreduction}{%
\begin{tikzpicture}[figure]
  \path[use as bounding box] (0,0) rectangle (1126,766);
  \draw[fill=figgray] (372,2) rectangle (705,71);
  \draw (372,12) -- (705,12) (413,2) -- (413,71);
  \node[figlabel=1.7] at (559,43) {selection gadgets};
  \foreach \figl/\figr in {3/137,181/316,473/608,651/786,943/1078}{
    \draw[fill=figgray, rounded corners=12pt] (\figl,260) rectangle (\figr,695);
  }
  \draw (538,71) -- (538,132);
  \foreach \figx/\figo in {40/e,96/h,219/e,275/h,511/e,567/h,689/e,744/h,981/e,1037/h}{
    \draw (538,132) -- (\figx,293);
    \figchain{\figx}{293}{39.6}{11}{\figo}
  }
  \draw[fill=white] (538,132) circle[radius=19];
  \node[figlabel=2.1] at (580,112) {$c$};
  \node[figlabel=2.1] at (395,460) {$\cdots$};
  \node[figlabel=2.1] at (865,460) {$\cdots$};
  \foreach \figx/\figt in {38/{1,2},217/{1,3},509/{1,p},685/{2,1}}{
    \node[figlabel=1.85] at (\figx,742) {$P_{\figt}^\edgel$};
    \node[figlabel=1.85] at (\figx+69,742) {$P_{\figt}^\heightl$};
  }
  \node[figlabel=1.85] at (952,742) {$P_{p,p-1}^\edgel$};
  \node[figlabel=1.85] at (1068,742) {$P_{p,p-1}^\heightl$};
\end{tikzpicture}%
}

\newcommand{\figbasemodel}{%
\begin{tikzpicture}[figure]
  \path[use as bounding box] (0,0) rectangle (238,960);
  \draw[fill=figgray, rounded corners=15pt] (1,85) rectangle (235,861);
  \draw (139,39) -- (139,145);
  \figchain{139}{145}{72.9}{17.5}{e}
  \draw[fill=white] (139,39) circle[radius=22];
  \node[figlabel] at (84,36) {$c$};
  \foreach \figk/\figq/\figt in {0/1/x,1/1/y,2/1/z,3/2/x,4/2/y,5/2/z,6/3/x,7/3/y,8/3/z}{
    \pgfmathsetmacro{\figy}{145+72.9*\figk}
    \node[figlabel] at (72,\figy) {$\figt_{i,j}^{\edgel,\figq}$};
    \draw[line width=4pt] (179,\figy-20) -- (179,\figy+20);
  }
  \node[figlabel] at (150,917) {$P_{i,j}^\edgel$};
\end{tikzpicture}%
}

\newcommand{\figrestrictionmodel}{%
\begin{tikzpicture}[figure]
  \path[use as bounding box] (0,0) rectangle (234,874);
  \draw[fill=figgray, rounded corners=15pt] (26,82) rectangle (232,774);
  \draw (84,36) -- (84,131);
  \figchain{84}{131}{59.6}{17}{e}
  \draw[fill=white] (84,36) circle[radius=22];
  \node[figlabel] at (30,34) {$c$};
  \draw[line width=4.5pt] (157,115) -- (157,680);
  \node[figlabel] at (165,722) {$\resmodel_{i,j}^\edgel$};
  \node[figlabel] at (88,828) {$P_{i,j}^\edgel$};
\end{tikzpicture}%
}

\newcommand{\figvertexmodel}{%
\begin{tikzpicture}[figure]
  \path[use as bounding box] (0,0) rectangle (466,874);
  \draw[fill=figgray, rounded corners=25pt] (26,82) rectangle (464,774);
  \draw (84,36) -- (84,131);
  \figchain{84}{131}{59.6}{17}{e}
  \draw[fill=white] (84,36) circle[radius=22];
  \node[figlabel] at (30,34) {$c$};
  \foreach \figx/\figy/\figq in {183/234/1,289/413/2,395/592/3}{
    \draw[model=figvertex, line width=4.5pt] (\figx,\figy) -- (\figx,680);
    \node[figlabel] at (\figx,722) {$\vermodel_{i,j}^{\edgel,\figq}$};
  }
  \node[figlabel] at (88,828) {$P_{i,j}^\edgel$};
\end{tikzpicture}%
}

\newcommand{\figedgemodel}{%
\begin{tikzpicture}[figure]
  \path[use as bounding box] (0,0) rectangle (570,878);
  \draw[fill=figgray, rounded corners=17pt] (3,170) rectangle (241,785);
  \draw[fill=figgray, rounded corners=17pt] (325,170) rectangle (563,785);
  \foreach \figx/\figo in {48/e,158/h,407/e,517/h}{
    \draw (283,42) -- (\figx,216);
    \figchain{\figx}{216}{55.5}{16}{\figo}
  }
  \draw[model=figedge] (283,64) -- (90,190) -- (90,286);
  \draw[model=figedge] (283,64) -- (196,190) -- (196,236);
  \draw[model=figedge] (283,64) -- (367,195) -- (367,622);
  \draw[model=figedge] (283,64) -- (474,195) -- (474,568);
  \draw[fill=white] (283,42) circle[radius=22];
  \node[figlabel] at (230,34) {$c$};
  \node[figlabel] at (54,830) {$P_{i,j}^\edgel$};
  \node[figlabel] at (165,830) {$P_{i,j}^\heightl$};
  \node[figlabel] at (407,830) {$P_{j,i}^\edgel$};
  \node[figlabel] at (517,830) {$P_{j,i}^\heightl$};
\end{tikzpicture}%
}

\newcommand{\figheightmodel}{%
\begin{tikzpicture}[figure]
  \path[use as bounding box] (0,0) rectangle (986,874);
  \foreach \figl in {3,325,731}{
    \draw[fill=figgray, rounded corners=17pt] (\figl,170) rectangle (\figl+238,785);
  }
  \foreach \figx/\figo in {48/e,158/h,407/e,517/h,813/e,923/h}{
    \draw (517,42) -- (\figx,216);
    \figchain{\figx}{216}{55.5}{16}{\figo}
  }
  \foreach \figx/\figy in {90/236,196/286,367/236,472/286,779/236,885/286}{
    \draw[model=figheight] (517,64) -- (\figx,190) -- (\figx,\figy);
  }
  \draw[fill=white] (517,42) circle[radius=22];
  \node[figlabel] at (462,34) {$c$};
  \node[figlabel] at (647,490) {$\cdots$};
  \node[figlabel] at (54,830) {$P_{i,1}^\edgel$};
  \node[figlabel] at (165,830) {$P_{i,1}^\heightl$};
  \node[figlabel] at (412,830) {$P_{i,2}^\edgel$};
  \node[figlabel] at (523,830) {$P_{i,2}^\heightl$};
  \node[figlabel] at (818,830) {$P_{i,p}^\edgel$};
  \node[figlabel] at (928,830) {$P_{i,p}^\heightl$};
\end{tikzpicture}%
}

\newcommand{\figselectiongadget}{%
\begin{tikzpicture}[gadget figure]
  \path[use as bounding box] (0,0) rectangle (1148,580);
  \figattached{232}{33}
  \draw (232,63) -- (232,166);
  \draw[fill=white] (232,196) circle[radius=30];
  \node[figlabel=3] at (318,200) {$c_\selgadl$};
  \foreach \figx in {34,161,413}{
    \draw (232,226) -- (\figx,378);
    \draw[fill=white] (\figx,408) circle[radius=30];
  }
  \foreach \figx in {79,207,465}{
    \draw[line width=4pt] (\figx,380) -- (\figx,437);
  }
  \node[figlabel=3] at (310,404) {$\cdots$};
  \draw[line width=.7pt, decorate, decoration={brace, mirror, amplitude=9pt}] (5,458) -- (466,458);
  \node[figlabel=3] at (235,538) {$\ell+1$};
  \figarrow{578}{270}
  \figattached{968}{172}
  \draw (968,202) -- (968,303);
  \figselection{792}{303}{1146}{407}
\end{tikzpicture}%
}

\newcommand{\figboundgadget}{%
\begin{tikzpicture}[gadget figure]
  \path[use as bounding box] (0,0) rectangle (1564,638);
  \figattached{440}{33}
  \draw (440,63) -- (440,165);
  \draw (440,226) -- (130,343) (440,226) -- (440,343) (440,226) -- (830,343);
  \draw[fill=white] (440,195) circle[radius=30];
  \node[figlabel=3] at (520,200) {$c_\bougadl$};
  \figselection{3}{343}{257}{417}
  \figselection{314}{343}{568}{417}
  \figselection{703}{343}{957}{417}
  \node[figlabel=3] at (635,376) {$\cdots$};
  \draw[line width=.7pt, decorate, decoration={brace, mirror, amplitude=20pt}] (5,462) -- (957,462);
  \node[figlabel=3] at (478,598) {$r$};
  \figarrow{1111}{270}
  \figattached{1472}{170}
  \draw (1472,200) -- (1472,304);
  \def\figboundlabel{$r$}
  \figbound{1385}{304}{1561}{378}{11}{3}
\end{tikzpicture}%
}

\newcommand{\figgroupedboundgadget}{%
\begin{tikzpicture}[figure]
  \path[use as bounding box] (0,0) rectangle (792,706);
  \draw[fill=white] (367,1) rectangle (401,35);
  \draw (384,35) -- (384,77);
  \draw (384,111) -- (131,202) (384,111) -- (384,202) (384,111) -- (692,202);
  \draw[fill=white] (384,94) circle[radius=17];
  \node[figlabel=2] at (430,97) {$c_\congadl$};
  \foreach \figx/\figl/\figs in {131/75/1,384/328/2,692/625/{\ell-1}}{
    \draw (\figx,202) -- (\figx,520);
    \draw (\figx,520) -- (\figx-66,598) (\figx,520) -- (\figx+26,651);
    \foreach \figy/\figq in {217/1,338/2,458/3}{
      \figdiamond{\figx}{\figy}{16}
      \node[figlabel=2] at (\figl,\figy) {$v_{\figs}^{\figq}$};
    }
    \foreach \figy/\figq in {280/1,400/2,520/3}{
      \figstar{\figx}{\figy}{17}
      \node[figlabel=2] at (\figl,\figy) {$w_{\figs}^{\figq}$};
    }
    \figselection{\figx-129}{597}{\figx-3}{632}
    \def\figboundlabel{$\ell-2$}
    \figbound{\figx-44}{651}{\figx+98}{704}{9}{2.1}
  }
  \node[figlabel=2.3] at (503,355) {$\cdots$};
\end{tikzpicture}%
}

\usepackage{makecell}
\usepackage{subcaption}
\usepackage{enumerate}
\usepackage{comment} 
\usepackage[numbered]{bookmark}

\newcommand{\rev}[1]{#1}

\newcommand{\natnum}{\mathbb{N}_{0}}
\newcommand{\natnump}{\mathbb{N}_{1}}

\newcommand{\mim}{\mathsf{mim}}
\newcommand{\dmax}{\mu}

\newcommand{\neio}[2]{N_{#1}(#2)}
\newcommand{\neic}[2]{N_{#1}[#2]}
\newcommand{\neior}[3]{N_{#1}^{#2}(#3)}

\newcommand{\degv}[2]{d_{#1}(#2)}
\newcommand{\maxdeg}[1]{\Delta(#1)}
\newcommand{\dist}[3]{\mathsf{dist}_{#1}(#2,#3)}

\newcommand{\lf}{\mathsf{lf}}
\newcommand{\Tmodel}{\mathcal{M}}
\newcommand{\selgad}{\mathscr{S}}

\newcommand{\bougad}{\mathscr{B}}
\newcommand{\congad}{\mathscr{G}}
\newcommand{\selgadl}{s}

\newcommand{\bougadl}{b}
\newcommand{\congadl}{g}
\newcommand{\edgel}{\mathsf{e}}
\newcommand{\heightl}{\mathsf{h}}
\newcommand{\bottoml}{\mathsf{b}}
\newcommand{\topl}{\mathsf{t}}
\newcommand{\numpart}{p}
\newcommand{\vermodel}{N}
\newcommand{\edgemodel}{L}
\newcommand{\himodel}{H}
\newcommand{\resmodel}{R}

\newcommand{\an}{\mathsf{an}}
\newcommand{\mem}[3]{\mathsf{mem}_{#1}(#2,#3)}
\newcommand{\memD}[2]{\mathsf{mem}_{#1}(#2)}
\newcommand{\problem}{\textsc{Min.\ $(\sigma, \rho)$-Dominating Set}}
\newcommand{\ecc}{\epsilon}
\newcommand{\tier}{V} %Tier
\newcommand{\DPsubG}[1]{G_{\le #1}} %Tier
\newcommand{\dpast}[3]{f_{#1}(#2,#3)}

\begin{document}

\maketitle

\begin{abstract}
The \textsc{Dominating Set} problem is known to be fixed-parameter tractable when parameterized by leafage, a structural graph parameter defined on chordal graphs.
We investigate whether this tractability extends to its variants with natural constraints. 
Specifically, we study the \textsc{Minimum Membership Dominating Set} problem and its generalization, the \textsc{Min.\ $(\sigma, \rho)$-Dominating Set} problem. 
Given a graph $G$ and a positive integer $\ell$, \textsc{Minimum Membership Dominating Set} asks whether there is a dominating set $D$ such that every vertex in $G$ is dominated by at most $\ell$ vertices in $D$.
We first show a negative result: \textsc{Minimum Membership Dominating Set} parameterized by leafage plus $\ell$ is W[1]-hard, demonstrating that even this natural constraint serves as a severe barrier to designing FPT algorithms. 
To complement this, we present an XP algorithm for the \textsc{Min.\ $(\sigma, \rho)$-Dominating Set} problem parameterized by asteroidal number plus $\mu$, where $\sigma$ and $\rho$ are finite sets of non-negative integers and $\mu$ is the maximum value in $\sigma \cup \rho$. 
Note that asteroidal number generalizes leafage to arbitrary graphs.
Combining this algorithm with our structural bound that every graph $G$ admits a dominating set such that no vertex is dominated by more than $7\an(G)-6$ vertices, where $\an(G)$ denotes the asteroidal number of $G$, we show that \textsc{Minimum Membership Dominating Set} admits an XP algorithm parameterized solely by asteroidal number.
\end{abstract}

\newpage

\section{Introduction}
A \emph{dominating set} of a graph $G$ is a subset $D\subseteq V(G)$ such that every vertex in $V(G)\setminus D$ is adjacent to some vertex in $D$.
In the \textsc{Dominating Set} problem, given a graph $G$ and a non-negative integer $k$, we are asked whether $G$ has a dominating set $D$ such that $|D| \le k$.
\textsc{Dominating Set} is one of the most intensively studied problems in the field of graph algorithms because the concept of dominating sets arises in a variety of practical applications, including network design, facility location, and error-correcting codes.

Although \textsc{Dominating Set} is NP-complete even for chordal graphs (that is, graphs without induced cycles of length at least~4), this computational hardness can be circumvented by exploiting \emph{leafage}, which is a structural graph parameter introduced in~\cite{LinMW98}.
The leafage of a chordal graph $G = (V, E)$ is defined as follows.
Consider a pair $(T, \Tmodel)$, where $T$ is a tree and $\Tmodel$ is a collection of subtrees of $T$.
To avoid confusion with the vertices of $G$, we refer to the vertices of $T$ as \emph{nodes} throughout this paper.
The pair $(T, \Tmodel)$ is called a \emph{tree representation} of $G$ if there is a bijection $f: V \to \Tmodel$ such that for any two distinct vertices $u, v \in V$, $uv \in E$ if and only if $f(u) \cap f(v) \neq \emptyset$ (that is, the subtrees $f(u)$ and $f(v)$ share at least one node in $T$).
Note that every chordal graph has a tree representation~\cite{Gavril74}, but such a representation is not necessarily unique.
The \emph{leafage} of $G$, denoted by $\lf(G)$, is defined as the minimum number of leaves in $T$ over all tree representations $(T,\Tmodel)$ of $G$~\cite{LinMW98}.

Recent studies have highlighted the power of leafage as a parameter to obtain FPT algorithms for problems that are intractable on chordal graphs.
Specifically, \textsc{Dominating Set}~\cite{FominGR20,GalbyMSST24}, \textsc{Connected Dominating Set}, and \textsc{Steiner Tree}~\cite{GalbyMSST24} admit such FPT algorithms, overcoming their NP-completeness~\cite{WhiteFP85}.
Moreover, Arvind et al.~\cite{ArvindNPZ26} established an FPT result for testing chordal graph isomorphism, a problem known to be GI-complete~\cite{ZemlyachenkoNR85}.
Since every interval graph has leafage at most~2, all these approaches serve as natural generalizations of algorithms for interval graphs.

For several problems, such as \textsc{Subset Feedback Vertex Set} and \textsc{Multicut with Undeletable Terminals}, XP algorithms parameterized by leafage are known~\cite{GalbyMSST24,PapadopoulosT24}.
On the negative side, however, both of these problems are W[1]-hard with respect to this parameter~\cite{GalbyMSST24,PapadopoulosT24}.

\subsection{Our contributions} \label{subsec:contributions}
While \textsc{Subset Feedback Vertex Set} and \textsc{Multicut with Undeletable Terminals} are W[1]-hard with respect to leafage, a fundamental question naturally arises: can the fixed-parameter tractability of \textsc{Dominating Set} be extended to variants with natural constraints?
To address this question, we investigate the computational complexity of the \textsc{Minimum Membership Dominating Set} problem.

For a vertex $v$ and a vertex subset $S$ of a graph $G=(V, E)$, the \emph{membership} $\mem{G}{v}{S}$ of $v$ is defined as \rev{$|\neic{G}{v} \cap S|$}. 
Let $\memD{G}{S} = \max_{v\in V} \mem{G}{v}{S}$.
Given a graph $G$ and a positive integer $\ell$, \textsc{Minimum Membership Dominating Set} determines whether $G$ has a dominating set $D$ such that $\memD{G}{D} \le \ell$.
In other words, it asks whether there is a dominating set $D$ such that every vertex in $G$ is dominated by at most $\ell$ vertices in $D$.
\textsc{Minimum Membership Dominating Set} has been researched for various graph parameters, such as treewidth, clique-width, and twin cover~\cite{AgrawalCNNR23,KarthikaMBBSS26,ReddyS24}. 

In this paper, we present the following negative result in response to our question.
\begin{restatable}{theorem}{thmleafage}
    \label{thm:leafage}
    \textsc{Minimum Membership Dominating Set} parameterized by leafage plus $\ell$ is W[1]-hard.
\end{restatable}
This result stands in sharp contrast to the original \textsc{Dominating Set}, which admits an FPT algorithm when parameterized by leafage, and demonstrates that even a natural constraint serves as a severe barrier to designing FPT algorithms.

To complement this negative result, we design an XP algorithm for \textsc{Minimum Membership Dominating Set} parameterized by leafage. 
More generally, for two finite sets $\sigma$ and $\rho$ of non-negative integers, we provide a comprehensive theorem applicable to \textsc{Min.\ $(\sigma,\rho)$-Dominating Set} parameterized by asteroidal number and the largest integer in $\sigma\cup\rho$.
% we provide a comprehensive theorem applicable to the \textsc{Min.\ $(\sigma, \rho)$-Dominating Set} problem parameterized by \emph{asteroidal number}.
% (The definitions of the problem and asteroidal number will be given in \Cref{sec:pre}.)

\begin{table}[t] 
    \caption{Partial list of problems defined by a $(\sigma, \rho)$-dominating set and their complexity when parameterized by leafage ($\lf$), asteroidal number ($\an$), and mim-width ($\mim$). 
    The notation ``W[1]-h'' stands for W[1]-hard.
    We write $\natnum = \{0,1,\ldots \}$, $\natnump = \{1,2,\ldots \}$, $\natnum^\ell = \{0,1,\ldots, \ell \}$, and $\natnump^\ell = \{1,2,\ldots, \ell \}$, where $\ell$ is a positive integer.
    For the XP results in the mim-width column, a branch decomposition is assumed to be given.
    Note that references to our results are omitted when they follow directly from results on other parameters.
    For instance, \Cref{thm:leafage} also implies the W[1]-hardness for \textsc{Min.\ Membership Dom.\ Set} 
    parameterized by $\an$ and $\mim$.
    }
    \label{tbl:results}
    \centering
    \begin{tabular}{|c|c|c||c|c|c|}
        \hline
        Problem name & $\sigma$ & $\rho$ & $\lf$ & $\an$ & $\mim$ \\
        \hline\hline
        \textsc{Dominating Set} &  $\natnum$ & $\natnump$ & FPT~\cite{FominGR20,GalbyMSST24} & XP~\cite{FominKM04} &  \makecell{XP~\cite{BuixuanTV13}\\ W[1]-h~\cite{FominGR20}} \\
        \hline
        \textsc{Total Dominating Set} & $\natnump$ & $\natnump$ & XP & XP~\cite{FominKM04} & \makecell{XP~\cite{BuixuanTV13}\\ W[1]-h~\cite{JaffkeKST18}} \\
        \hline
        \textsc{Efficient Dom.\ Set} & $\{0 \}$ & $\{1 \}$ & XP &  XP~\cite{BroersmaKKM99} & XP~\cite{BuixuanTV13} \\
        \hline
        \textsc{Total Perfect Dom.\ Set} & $\{1 \}$ & $\{1 \}$ & XP & XP~[Th.~\ref{thm:domset_bound}] & XP~\cite{BuixuanTV13} \\
        \hline
        \textsc{Weakly Perfect Dom.\ Set} & $\{0,1 \}$ & $\{1 \}$ & XP & XP~[Th.~\ref{thm:domset_bound}] & XP~\cite{BuixuanTV13} \\
        \hline
        \textsc{Min.\ Membership Dom.\ Set} & $\natnum^{\ell-1}$ & $\natnump^\ell$ & \makecell{XP\\W[1]-h~[Th.~\ref{thm:leafage}]}  & \makecell{XP~[Th.~\ref{thm:fair_member}]\\W[1]-h} & W[1]-h \\
        \hline
    \end{tabular}
\end{table}

\rev{Let $\sigma$ and $\rho$ be (possibly infinite) subsets of non-negative integers.
A vertex subset $D$ of a graph $G = (V, E)$ is a \emph{$(\sigma, \rho)$-dominating set} if the following two conditions are satisfied: 
\begin{enumerate}
    \item$|\neio{G}{v} \cap D| \in \sigma$ for every vertex $v \in D$; and 
    \item $|\neio{G}{v} \cap D| \in \rho$ for every vertex $v \in V \setminus D$, 
\end{enumerate}
where $\neio{G}{v}$ denotes the open neighborhood of $v$ in $G$.
Given a graph $G$ and a non-negative integer $k$, the \textsc{Min.\ $(\sigma, \rho)$-Dominating Set} problem aims to determine whether $G$ has a $(\sigma, \rho)$-dominating set $D$ such that $|D| \le k$.
\textsc{Min.\ $(\sigma, \rho)$-Dominating Set} can represent \textsc{Dominating Set} and a wide range of its variants, as shown in \Cref{tbl:results}.
For example, \textsc{Dominating Set} and \textsc{Total Dominating Set} are equivalent to \textsc{Min.\ $(\natnum, \natnump)$-Dominating Set} and \textsc{Min.\ $(\natnump, \natnump)$-Dominating Set}, respectively, where $\natnum = \{0,1,\ldots \}$ and $\natnump = \{1,2,\ldots \}$.
% Moreover, given the upper bound $\ell$ for $\sigma$ and $\rho$ as part of the input, \textsc{Min.\ $(\{0,\ldots, \ell-1\}, \{1,\ldots, \ell\})$-Dominating Set} generalizes \textsc{Minimum Membership Dominating Set}.}
Throughout this paper, we assume that both $\sigma$ and $\rho$ are non-empty, since the cases where either set is empty can be handled directly.}

\rev{A vertex subset $I$ of $G$ is \emph{independent} if any two vertices in $I$ are not adjacent.
An independent set $A$ of a graph $G$ is an \emph{asteroidal set} if for each $a \in A$, all vertices in $A \setminus \{a\}$ are contained in one connected component of $G - \neic{G}{a}$.
The \emph{asteroidal number} of $G$, denoted by $\an(G)$, is the maximum size of an asteroidal set in $G$.
(See \Cref{fig:asteroidal} as an example.)
Intuitively, the asteroidal number measures the ``linearity'' of a graph. 
Graphs with a large asteroidal number radiate outward from certain parts (such as the three vertices $b_1$, $b_2$, and $b_3$ in \Cref{fig:asteroidal}(a)), and are thus far from having a linear structure.}

\begin{figure}[t]
    \begin{minipage}{0.45\textwidth}
        \centering
        \begin{tikzpicture}[every node/.style={circle, draw, very thick, inner sep=1pt,}, every path/.style={very thick}]
              \node (b1) at (90:0.75) {$b_1$};
              \node (b2) at (210:0.75) {$b_2$};
              \node (b3) at (330:0.75) {$b_3$};
            
              \node (a1) at ($(b1)+(90:1)$) {$a_1$};
              \node (a2) at ($(b2)+(-150:1)$) {$a_2$};
              \node (a3) at ($(b3)+(-30:1)$) {$a_3$};
            
              \draw (b1) -- (b2) -- (b3) -- (b1); 
              \draw (a1) -- (b1);
              \draw (a2) -- (b2);
              \draw (a3) -- (b3);
        \end{tikzpicture} \\
        (a)
    \end{minipage}
    \begin{minipage}{0.45\textwidth}
        \centering
        \begin{tikzpicture}[every node/.style={circle, draw, very thick, inner sep=1pt,}, every path/.style={very thick}]
              \node (b1) at (90:0.75) {$b_1$};
              \node (b2) at (210:0.75) {$b_2$};
              \node (b3) at (330:0.75) {$b_3$};
            
              \node (a1) at ($(b1)+(90:1)$) {$a_1$};
              \node (a2) at ($(b2)+(-150:1)$) {$a_2$};
              \node (a3) at ($(b3)+(-30:1)$) {$a_3$};
            
              \draw (b2) -- (b1) -- (b3); 
              \draw[->] (b2) -- (b3); 
              \draw (a1) -- (b1);
              \draw[->] (a2) -- (b2);
              \draw[<-] (a3) -- (b3);
        \end{tikzpicture} \\
        (b)
    \end{minipage}
    \caption{(a) A graph $G$ with an asteroidal triple $\{a_1, a_2, a_3\}$. As shown by directed edges in~(b), there is a path between $a_2$ and $a_3$ that avoids $\neic{G}{a_1} = \{a_1, b_1\}$. Similarly, there is a path between $a_3$ and $a_1$ (resp.\ $a_1$ and $a_2$) that avoids  $\neic{G}{a_2}$ (resp.\ $\neic{G}{a_3}$).
    Therefore, the set $\{a_1, a_2, a_3\}$ is an asteroidal triple of $G$.}
    \label{fig:asteroidal}
\end{figure}
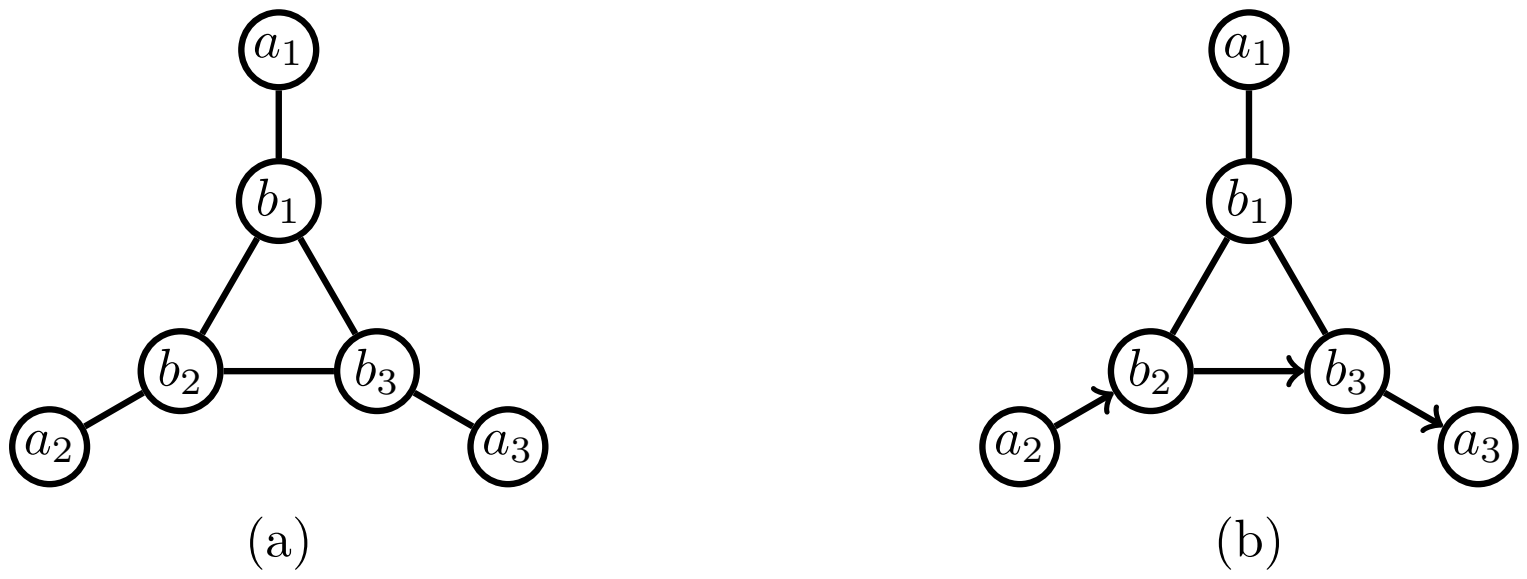

\begin{figure}[t]
	\centering
	\begin{tikzpicture}
		\newcommand{\width}{3}
		\newcommand{\height}{1.25}
		\node at (0, 0)[draw, rectangle, very thick] (an){bounded asteroidal number};
        \node at (0, -1*\height)[draw, rectangle, very thick, fill = white] (AT){AT-free};
        \node at (1.5*\width, -\height)[draw, rectangle, very thick] (IS){bounded independence number};
        \node at (-1.25*\width, -2*\height)[draw, rectangle, very thick] (lf){bounded leafage};
        \node at (0, -2*\height)[draw, rectangle, very thick, fill = white] (co-comparability){co-comparability};
        \node at (-1.25*\width, -3*\height)[draw, rectangle, very thick] (interval){interval};
		\node at (0*\width, -3*\height)[draw, rectangle, very thick] (permutation){permutation};
        \node at (1*\width, -3*\height)[draw, rectangle, very thick] (co-bipartite){co-bipartite};
		\node at (2*\width, -3*\height)[draw, rectangle, very thick] (co-planar){co-planar};

		\draw [very thick, ->] (an) -- (AT);
        \draw [very thick, ->] (an) -- (IS);
        \draw [very thick, ->] (an) -- (lf);

        \draw [very thick, ->] (lf) -- (interval);
        \draw [very thick, ->] (AT) -- (co-comparability);
        \draw [very thick, ->] (IS) -- (co-bipartite);
        \draw [very thick, ->] (IS) -- (co-planar);

		\draw [very thick, ->] (co-comparability) -- (co-bipartite);
        \draw [very thick, ->] (co-comparability) -- (interval);
        \draw [very thick, ->] (co-comparability) -- (permutation);

	\end{tikzpicture}
	\caption{Graph classes that have bounded asteroidal number.
    Each arrow represents the inclusion relationship between graph classes: 
	$A \rightarrow B$ means that the graph class $B$ is a subclass of the graph class $A$.}
	\label{fig:graphclass}
\end{figure}
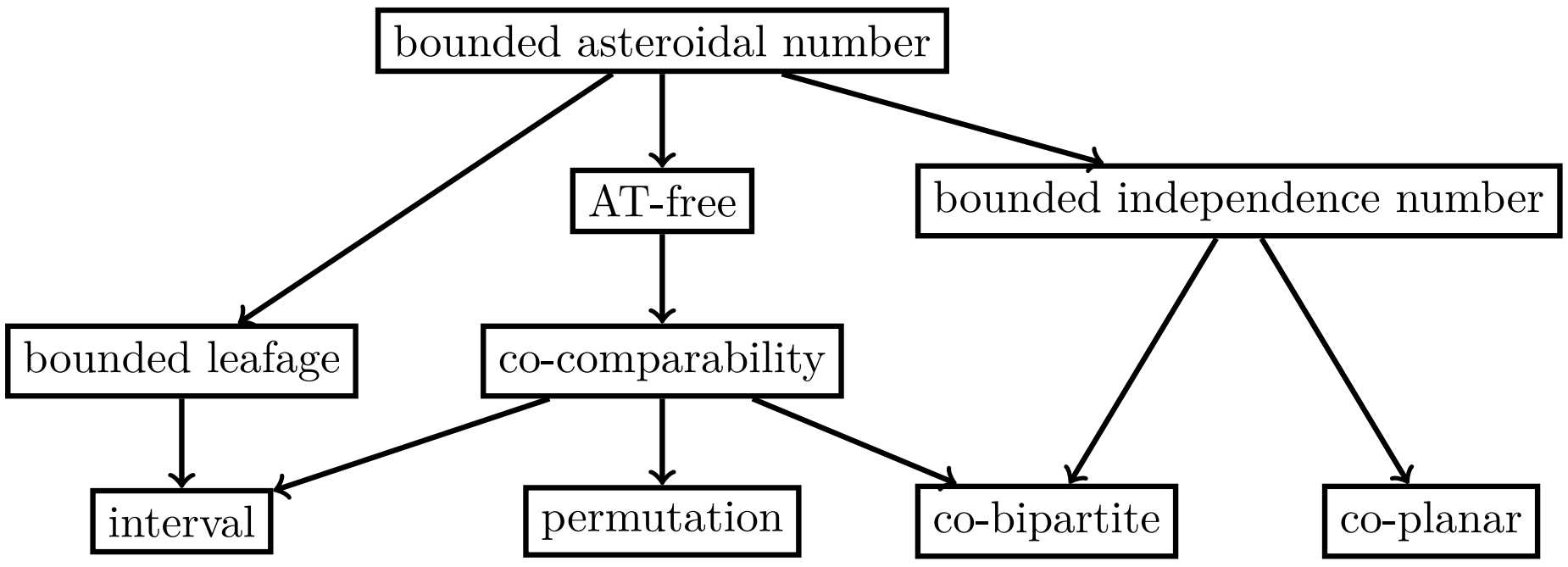

\rev{A graph $G$ has asteroidal number~$1$ if and only if $G$ is a complete graph.  
Every interval graph and permutation graph has asteroidal number at most~$2$.
Moreover, graphs with asteroidal number at most~$2$ are exactly the \emph{asteroidal triple-free} (\emph{AT-free}) graphs.
AT-free graphs have been the subject of extensive research because they provide a generalization of several well-known graph classes, including interval graphs, permutation graphs, and co-comparability graphs. 
Furthermore, graphs of bounded asteroidal number generalize graphs of bounded independence number and graphs of bounded leafage (see \cref{fig:graphclass}).
It is known that for any chordal graph $G$, we have $\an(G) \le \lf(G)$~\cite{LinMW98}.
Note that while leafage is defined only for chordal graphs, the asteroidal number serves as a generalization of this parameter to arbitrary graphs.}

For finite subsets $\sigma$ and $\rho$ of $\natnum$, we denote $\sigma_{\max} = \max \sigma $ and $\rho_{\max} = \max \rho$.
We give the following theorem, which demonstrates the polynomial-time solvability of many variants of \textsc{Dominating Set}.

\begin{restatable}{theorem}{thmdomsetbound}
\label{thm:domset_bound}
    Let $\sigma$ and $\rho$ be finite subsets of $\natnum$, and let $\dmax = \max\{\sigma_{\max}, \rho_{\max} \}$.
    Then there exists an XP algorithm for \textsc{Min.\ $(\sigma, \rho)$-Dominating Set} parameterized by asteroidal number plus $\dmax$.
\end{restatable}

Crucially, \textsc{Min.\ $(\{0,\ldots, \ell-1\}, \{1,\ldots, \ell\})$-Dominating Set} generalizes \textsc{Minimum Membership Dominating Set}, and graphs of bounded asteroidal number contain various graph classes, including graphs of bounded leafage~\cite{LinMW98}.
Since $\dmax = \ell$ for \textsc{Min.\ $(\{0,\ldots, \ell-1\}, \{1,\ldots, \ell\})$-Dominating Set}, \Cref{thm:domset_bound} directly yields an XP algorithm for \textsc{Minimum Membership Dominating Set} when parameterized by the asteroidal number plus $\ell$, providing \rev{an algorithmic counterpart} to the negative result of \Cref{thm:leafage}.

To further strengthen this result for \textsc{Minimum Membership Dominating Set}, we employ the following theorem.

\begin{restatable}{theorem}{thmdomOPT} 
\label{thm:dom_OPT}
    Every graph $G$ has a dominating set $D$ such that $\memD{G}{D} \le 7\an(G)-6$.
\end{restatable}

\Cref{thm:domset_bound} and \Cref{thm:dom_OPT} establish the following theorem.

\begin{restatable}{theorem}{thmfairmember} 
\label{thm:fair_member}
    There exists an XP algorithm for \textsc{Minimum Membership Dominating Set} parameterized by asteroidal number.
\end{restatable}

\Cref{tbl:results} summarizes our results and known results with respect to leafage, asteroidal number, and mim-width, which is a graph parameter known to be bounded by leafage.
We will also briefly discuss how our tractable result can be extended to the distance-$r$ variants for a fixed integer $r \in \natnump$.

\subparagraph{Future work.}
As summarized in \Cref{tbl:results}, the fixed-parameter tractability of several variants of \textsc{Dominating Set}, such as \textsc{Total Dominating Set} and \textsc{Efficient Dominating Set}, with respect to leafage remains open. 
Establishing whether these problems are fixed-parameter tractable or W[1]-hard is a natural direction for future work. 
Furthermore, while we can show that \Cref{thm:dom_OPT} extends to graphs admitting a $d$-octopus (a structural generalization of asteroidal number), it remains open whether the algorithmic result of \Cref{thm:domset_bound} similarly applies to this parameter. 
Investigating this generalization would provide deeper insights into the tractability of these problems.

\subsection{Technical overview}

\subparagraph{W[1]-hardness.}
To prove \Cref{thm:leafage}, we provide an FPT reduction from \textsc{Multicolored Clique}, which is known to be W[1]-hard when parameterized by the solution size $\numpart$~\cite{FellowsHRV09}.
Since every chordal graph admits a tree representation, rather than constructing the graph itself, we build a tree representation where the number of leaves is bounded solely by a function of $p$.

The key ingredient in our construction is a star with the center $c$ and $\ell+1$ leaves. 
If $c$ is not included in a solution $D$ for \textsc{Minimum Membership Dominating Set}, then all $\ell+1$ leaves must belong to $D$ to be dominated. 
However, this implies that $c$ has $\ell+1$ neighbors in $D$, which contradicts $\memD{G}{D} \le \ell$. 
Thus, we conclude that $c \in D$.
Moreover, repeated use of this structure allows us to designate vertices that are excluded from the solution $D$.

To those experienced in complexity reductions, establishing intractability might seem relatively easy if one can enforce these two properties on several vertices. 
The true challenge in this reduction, however, lies in bounding the membership of all vertices by $\ell$ while restricting the number of leaves in the tree representation by a function of $p$. 
To accomplish this, we must construct a series of intricate gadgets.

\subparagraph{XP algorithms.}
To prove \Cref{thm:domset_bound}, we employ the technique of BFS layers, a method previously used to establish the tractability of \textsc{Dominating Set} on asteroidal triple-free graphs~\cite{Kratsch00} and graphs with a $d$-octopus~\cite{FominKM04}.
% Given a graph $G$, arbitrarily choose a vertex $s$. 
% For an integer $i$, let $\tier_i = \{ v \in V(G) : \dist{G}{s}{v} = i \}$.
% As shown in \cite{FominKM04}, for any pair of integers $i$ and $j$, there exists a minimum dominating set $D$ of $G$ such that the size of $D \cap \bigcup_{s=i}^{i+j} \tier_s$ is bounded by $(j+5)d-1$.
% This property allows us to employ a dynamic programming approach.
Given a graph $G$, choose a vertex $s$ and let $V_i=\{v\in V(G):\operatorname{dist}_G(s,v)=i\}$ for each non-negative integer $i$. 
Suppose that $G$ has a dominating set $S$ such that \rev{$|S\cap V_i|\le d$} for every $i$, where $d$ is a positive integer. 
As shown in \cite{FominKM04}, there exists a minimum dominating set $D$ of $G$ such that, for every non-negative integer $j$, any $j+1$ consecutive BFS layers contain at most $(j+5)d-1$ vertices of $D$. This property allows us to employ a dynamic programming approach.

The preceding result for \textsc{Dominating Set} is proved by an exchange argument that transforms an arbitrary minimum dominating set into one satisfying the required bounds on consecutive BFS layers. 
However, these exchanges need not preserve the membership bound in \textsc{Minimum Membership Dominating Set} or the neighborhood constraints in \textsc{Min.\ $(\sigma,\rho)$-Dominating Set}. 
Thus, this argument does not directly extend to these problems. 
Instead, for finite sets $\sigma$ and $\rho$, we take a different route to establish that every $(\sigma,\rho)$-dominating set intersects each BFS layer in a number of vertices bounded by a function of $\an(G)$ and $\mu$.

\subsection{Related results}
The study of AT-free graphs was initiated by Lekkerkerker and Boland in 1962 to characterize interval graphs~\cite{LekkerkerkerB62}.
It is known that many problems that are NP-hard on general graphs become tractable on AT-free graphs; for example, \textsc{Independent Set}~\cite{BroersmaKKM99}, \textsc{Dominating Set}~\cite{Kratsch00}, \textsc{Total Dominating Set}~\cite{Kratsch00} and so on~\cite{BalakrishnanRR93,Chang03,GolovachPL22,KloksP21,KratschMT08,MukherjeeS23,Stacho12,TripathiKPPW22}.

In 1998, Broersma et al.\ introduced asteroidal number to generalize AT-free graphs~\cite{BroersmaKKM98}.
Graphs of bounded asteroidal number have received less attention than AT-free graphs. 
Known XP algorithms parameterized by asteroidal number include those for \textsc{Independent Set}, 
\textsc{Independent Dominating Set}, \textsc{Independent Perfect Dominating Set} (equivalently \textsc{Efficient Dominating Set})~\cite{BroersmaKKM99}, and \textsc{$k$-Coloring} for fixed $k$~\cite{KratschM12}.

Fomin et al.~\cite{FominKM04} studied graphs admitting a $d$-octopus.
They gave XP algorithms for \textsc{Dominating Set} and \textsc{Total Dominating Set} parameterized by $d$.
Since every connected graph of asteroidal number at most $a$ admits an $(a-1)$-octopus for $a\ge 2$, these results also yield XP algorithms parameterized by asteroidal number.

Various problems expressible in the form of \textsc{Min.\ $(\sigma, \rho)$-Dominating Set} are NP-complete for general graphs, whereas they are typically solvable in polynomial time if an interval graph or a permutation graph is given.
% This tractability was extended by \rev{Bui-Xuan} et al.\ as follows: when each of $\sigma$ and $\rho$ is finite or co-finite, \textsc{Min.\ $(\sigma, \rho)$-Dominating Set} admits an XP algorithm parameterized by mim-width~\cite{BuixuanTV13}, where mim-width is a structural graph parameter introduced by Vatshelle~\cite{Vatshelle12}.
More generally, when each of $\sigma$ and $\rho$ is finite or co-finite, \textsc{Min.\ $(\sigma,\rho)$-Dominating Set} admits an XP algorithm parameterized by $w$ when a branch decomposition of mim-width $w$ is given~\cite{BuixuanTV13,BelmonteV13}.
Here, mim-width is a structural graph parameter introduced by Vatshelle~\cite{Vatshelle12}.
For every chordal graph $G$, its mim-width $\mim(G)$ satisfies $\mim(G) \le \lf(G)$~\cite{FominGR20,PapadopoulosT24}.
It was shown in~\cite{BelmonteV13,Vatshelle12} that the following graphs have bounded mim-width: interval graphs, permutation graphs, distance-hereditary graphs, $k$-polygon graphs, circular $k$-trapezoid graphs, bounded treewidth graphs, bounded clique-width graphs, and so on.
Consequently, various problems are solvable in polynomial time for these graphs.
(More details are summarized in~\cite{JaffkeKST19}.)

Although \textsc{Minimum Membership Dominating Set} is also expressible in the form of \textsc{Min.\ $(\sigma, \rho)$-Dominating Set}, the sets $\sigma$ and $\rho$ depend on the input $\ell$. 
% Due to this, it remains open whether the problem admits an XP algorithm parameterized solely by mim-width, even though we can obtain an XP algorithm parameterized by mim-width and $\ell$.
Given a branch decomposition of mim-width $w$, the same framework yields an XP algorithm parameterized by $w+\ell$.
It remains open whether the problem admits an XP algorithm parameterized by $w$ alone, even when such a decomposition is given.

Furthermore, a major drawback of mim-width is that it is difficult to compute.
It was shown that for some constant $c$, the problem of determining whether a graph has mim-width at most $c$ is NP-complete~\cite{BergougnouxBD25}.
Recently, the problem was reported to remain NP-complete even for $c=2$~\cite{TourLN25}.
On the other hand, a straightforward XP algorithm exists to compute the asteroidal number of a given graph. 

Note that asteroidal number and mim-width are incomparable parameters. 
It is known that co-comparability graphs \rev{(which have asteroidal number at most~$2$)} have unbounded mim-width~\cite{KangKST17}, and it can be observed that trees (which have mim-width at most~$1$) have unbounded asteroidal number.
Consequently, approaches designed for bounded mim-width graphs cannot generally be applied to AT-free graphs and graphs with bounded asteroidal number.

\section{Preliminaries} \label{sec:pre}

For a positive integer $i$, we use $[i]$ as shorthand for the set $\{1, 2, \dots, i\}$ of integers.
We denote $\natnum = \{0,1,\ldots \}$ and $\natnump = \{1,2,\ldots \}$.
We also write $\natnum^{p} = \{0,\ldots,p\}$ and $\natnump^{p} = \{1,\ldots,p\}$ for a fixed integer $p\in \natnum$. 
When $p=0$, we define $\natnump^{p} = \emptyset$.

Let $G = (V, E)$ be a graph.
All graphs in this paper are simple, undirected, and unweighted.
We denote by $V(G)$ and $E(G)$ the vertex and edge sets of $G$, respectively. 
For a vertex subset $V' \subseteq V$, we denote by $G[V']$ the subgraph induced by $V'$.
We use $G-V'$ as shorthand for $G[V\setminus V']$.
For a vertex $v$ of $G$, we denote by $\neio{G}{v}$ and $\neic{G}{v}$ the \emph{open neighborhood} and \emph{closed neighborhood} of $v$ in $G$, respectively, that is, $\neio{G}{v} = \{ w \in V : vw \in E \}$ and $\neic{G}{v} = \neio{G}{v} \cup \{v\}$.
The \emph{degree} of a vertex $v$ of $G$, denoted by $\degv{G}{v}$, is the size of $\neio{G}{v}$. 
Let $\maxdeg{G}$ be the maximum degree of $G$.
A sequence $\langle v_0, v_1, \ldots, v_\ell \rangle$ of vertices in $G$ is called a $v_0$-$v_\ell$ \emph{path} if $v_{i-1}v_i \in E$ for each $i \in [\ell]$ and all vertices in the sequence are distinct.
The value $\ell$ is called the \emph{length} of the path.
A graph $G$ is \emph{connected} if for every pair of vertices $u,v \in V$, there exists a $u$-$v$ path of $G$. 
A \emph{connected component} of $G$ is a maximal connected subgraph of $G$.
The \emph{distance} between vertices $u$ and $v$ of $G$, denoted by $\dist{G}{u}{v}$, is defined as the length of a shortest $u$-$v$ path of $G$. 
If $u$ and $v$ belong to different connected components, then let $\dist{G}{u}{v} = +\infty$.

Throughout this paper, we assume that the input graph $G$ is connected.
Otherwise, consider the problem for each connected component of $G$.

\section{W[1]-hardness} \label{sec:hardness}

This section shows the W[1]-hardness of \textsc{Minimum Membership Dominating Set}, as stated in \Cref{thm:leafage}, by providing an FPT reduction from \textsc{Multicolored Clique}. 
\begin{itembox}[l]{\textsc{Multicolored Clique}}
    \begin{description}
        \item[Given:] A graph $G'=(V,E)$, an integer $\numpart\ge 2$, and a partition $(V_1, V_2, \ldots, V_\numpart)$ of $V$.
        \item[Task:] Determine whether there exists a clique $C$ of $G'$ such that $|C \cap V_i| =1$ for each $i\in [\numpart]$.
    \end{description}
\end{itembox}
\textsc{Multicolored Clique} is known to be W[1]-hard when parameterized by $\numpart$~\cite{FellowsHRV09}.
Without loss of generality, assume that $V_i$ is an independent set of $G'$ and $|V_1| = \dots = |V_\numpart| = n$.
Let $V_i  = \{v_i^1, v_i^2, \dots, v_i^n \}$ for each $i \in [\numpart]$.
For two integers $i, j \in [\numpart]$ with $i < j$, we denote by $E_{i,j}$ the set of edges joining a vertex in $V_i$ and a vertex in $V_j$.
We may assume that $E_{i,j} \neq \emptyset$ for all such $i$ and $j$, since otherwise the instance is a no-instance.

\subsection{High-level idea} \label{subsec:idea}
\begin{figure}[t]
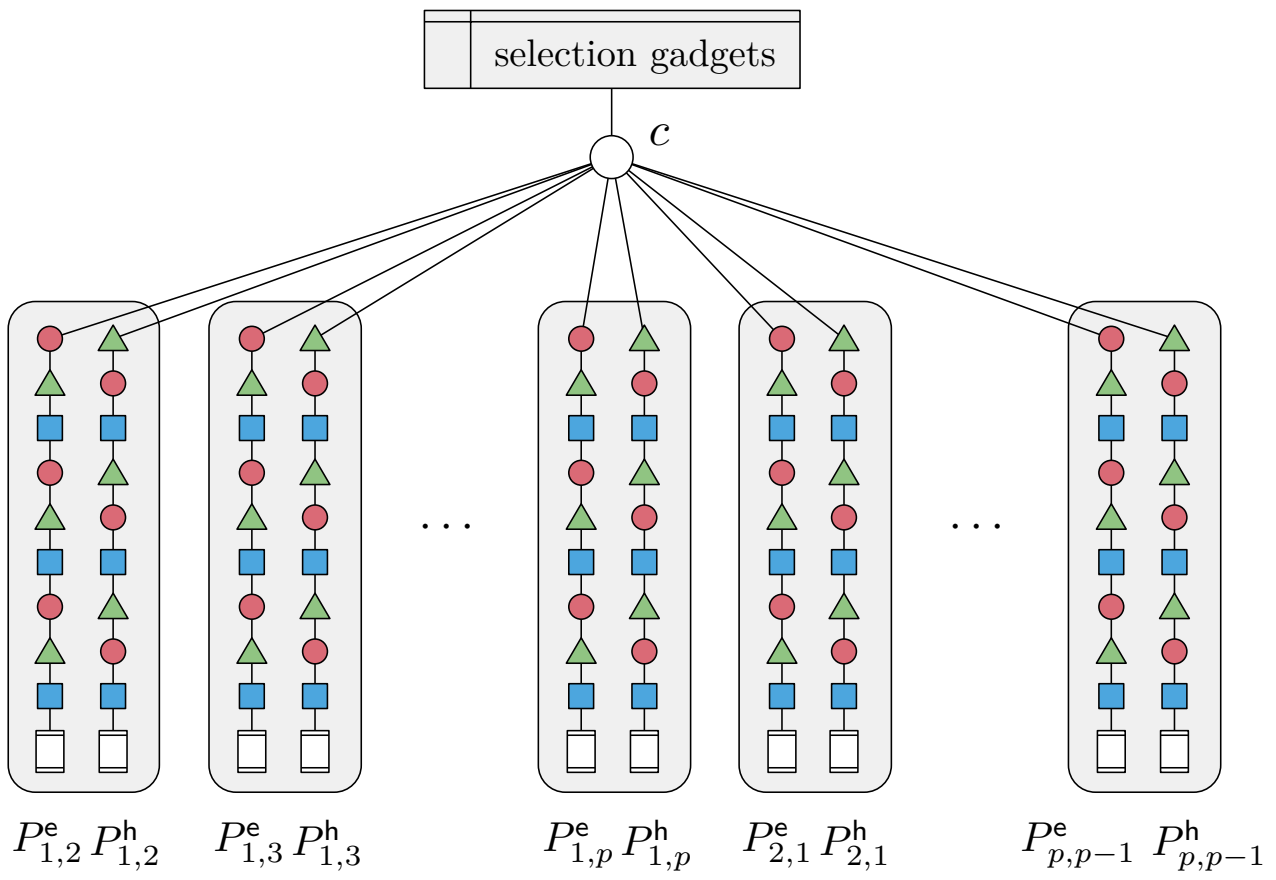

    \centering
    \resizebox{0.60\linewidth}{!}{\figreduction}
    \caption{An overview of the tree $T$ constructed by our reduction. The symbols at the bottom of this figure represent collections of selection gadgets, bound gadgets, and grouped bound gadgets.}
    \label{fig:reduction}
\end{figure}

Here, we provide an overview of our reduction.
We will construct an instance $(G,\ell)$ of \textsc{Minimum Membership Dominating Set} from an instance $(G',\numpart,(V_1, V_2, \ldots, V_\numpart))$ of \textsc{Multicolored Clique}, where $\ell = O(\numpart^2)$.
Instead of describing the construction of $G$, we will explain how to construct its tree representation $(T,\Tmodel)$.
Furthermore, we view \textsc{Minimum Membership Dominating Set} as the following equivalent problem on a tree representation.
For a model $M \in \Tmodel$ and a set $\Tmodel' \subseteq \Tmodel$, denote $\mem{T}{M}{\Tmodel'} = | \{ M' \in \Tmodel'  : M \cap M' \neq \emptyset \} |$.
Given a tree representation $(T,\Tmodel)$ of a chordal graph $G$ and a positive integer $\ell$, the objective is to determine whether there exists a set $ \Tmodel^\ast \subseteq \Tmodel$ that satisfies $\mem{T}{M}{\Tmodel^\ast} \in \natnump^{\ell}$ for every model $M \in  \Tmodel$.
Note that the two conditions originally imposed on a $(\sigma, \rho)$-dominating set can be expressed as a single condition as above.
In the following, when we refer to \textsc{Minimum Membership Dominating Set}, consider the problem on a tree representation $(T,\Tmodel)$.

\Cref{fig:reduction} illustrates the overall structure of $T$.
While some gadgets have been omitted, the core construction comprises a central node and paths emanating from it. 
A gadget is attached to the bottom of each path, which will be explained later.
We add models to $T$ representing the vertices and edges in $G'$.
Specifically, \emph{base models}, \emph{vertex models}, and \emph{edge models} form the core of the construction of $(T,\Tmodel)$. 
Each node $t$ of $T$ has a base model consisting solely of $t$ (\cref{fig:models}(a)).
Each blue square $z_{i,j}^{\edgel,q}$ corresponds to a vertex model $\vermodel_{i,j}^{\edgel,q}$, which extends from the bottom of the path $P_{i,j}^\edgel$ to $z_{i,j}^{\edgel,q}$ (\cref{fig:models}(c)).
As the name suggests, the vertex models correspond to the vertices of $G'$.
If $\vermodel_{i,j}^{\edgel,q}$ is included in a solution to \textsc{Minimum Membership Dominating Set}, then the base model on the green triangle $y_{i,j}^{\edgel,q}$, located immediately above $z_{i,j}^{\edgel,q}$, must be dominated by other models.
This role is fulfilled by an edge model, representing an edge in $E_{i,j}$ (\cref{fig:models}(d)).
Since this property holds for every path $P_{i,j}^\edgel$, the edge set corresponding to these edge models, which dominate the base models on the green triangles, forms a clique in $G'$.

This is an underlying idea of our reduction.
However, this alone is far from sufficient to make the reduction work. 
In what follows, we formally describe the construction of $(T,\Tmodel)$.

\begin{figure}[t]
    \centering
    \begin{tabular}{ccc}
    
        % T1
        \begin{minipage}[t]{0.3\linewidth}
          \centering
          \resizebox{0.3\linewidth}{!}{\figbasemodel}
        \end{minipage} &
        
        % T2
        \begin{minipage}[t]{0.3\linewidth}
          \centering
          \resizebox{0.3\linewidth}{!}{\figrestrictionmodel}
        \end{minipage} &
        
        % T3
        \begin{minipage}[t]{0.3\linewidth}
          \centering
          \resizebox{0.6\linewidth}{!}{\figvertexmodel}
        \end{minipage} \\

        (a) & (b) & (c) \\

 & & \\

        % T1
        \begin{minipage}[t]{0.3\linewidth}
          \centering
          \resizebox{0.75\linewidth}{!}{\figedgemodel}
        \end{minipage}
        &
        % T2
        % \begin{minipage}[t]{0.3\linewidth}
        % \end{minipage} &
        
        % T3
        \begin{minipage}[t]{0.3\linewidth}
          \centering
          \resizebox{1.3\linewidth}{!}{\figheightmodel}
        \end{minipage}
        & 

        \\

        (d) & (e) & 

    \end{tabular}
    \caption{This figure depicts models in bold lines: (a) the base models on $P_{i,j}^\edgel$; (b) the restriction model $\resmodel_{i,j}^\edgel$; (c) the vertex models $\vermodel_{i,j}^{\edgel,1}$, $\vermodel_{i,j}^{\edgel,2}$, and $\vermodel_{i,j}^{\edgel,3}$; (d) the edge model $\edgemodel_{i,j}^{1,3}$; and (e) the height model $\himodel_{i}^{1}$.}
    \label{fig:models}
  \end{figure}

\subsection{Gadgets} \label{subsec:gadgets}
In this subsection, we define three gadgets: a \emph{selection gadget}, a \emph{bound gadget}, and a \emph{grouped bound gadget}.
Additionally, we discuss the useful properties of the gadgets.
Note that each gadget has a specific node called the \emph{center} of the gadget.
% For a tree representation $(T, \Tmodel)$, a gadget is \emph{attached} to a node $t$ in $T$ if the center of the gadget is connected to $t$ by an edge, and adds the models defined by the gadget to $\Tmodel$.
For a tree representation $(T, \Tmodel)$, we \emph{attach} a gadget to a node $t$ of $T$ by connecting its center to $t$ with an edge and adding its models to $\Tmodel$.

\subparagraph{Selection gadget.}

\begin{figure}[t]
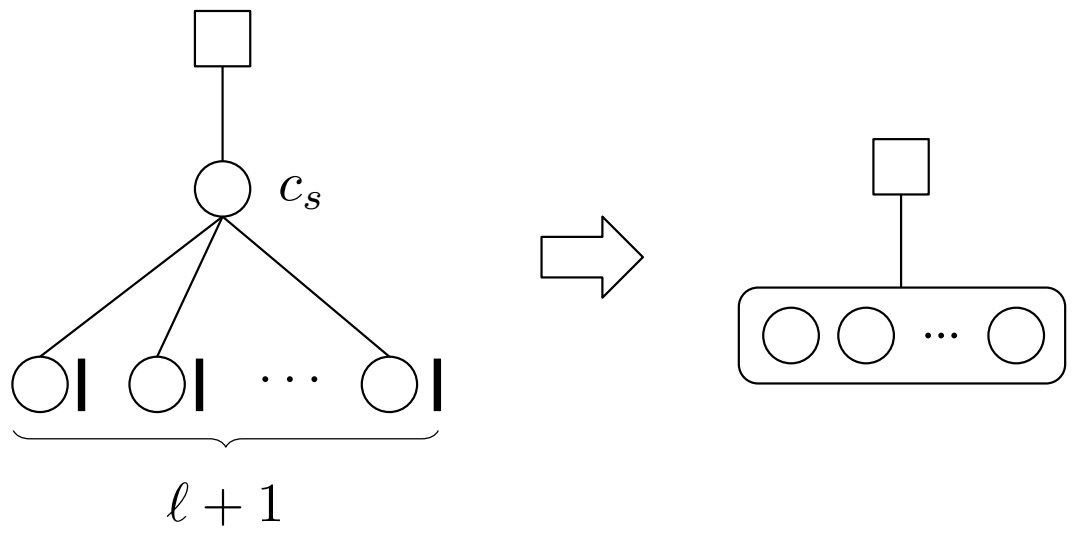

    \centering
    \resizebox{0.5\linewidth}{!}{\figselectiongadget}
    \caption{The figure on the left illustrates the selection gadget $\selgad$, where the square denotes a node to which the gadget is attached. For simplicity, the selection gadget is represented as shown in the figure on the right.}
    \label{fig:selection_gadget}
\end{figure}

A selection gadget $\selgad = (T_\selgadl,\Tmodel_\selgadl)$ is defined as follows.
(See also \Cref{fig:selection_gadget}.)
The tree $T_\selgadl$ consists of a unique internal node $c_\selgadl$ and $\ell+1$ leaves.
We call $c_\selgadl$ the \emph{center} of $T_\selgadl$.
For each node $w$ in $T_\selgadl$ except for $c_\selgadl$, we introduce a model consisting solely of $w$ to $\Tmodel_\selgadl$.
We call the models in $\Tmodel_\selgadl$ \emph{selection models}.
We say that a model $M$ is \emph{plugged into} $\selgad$ if $M$ intersects with all models in $\selgad$.
% The collection $\Tmodel_\selgadl$ is referred to as the \emph{models} for $\selgad$.

% The selection gadget $\selgad$ will be used to construct the resulting tree representation $(T,\Tmodel)$ and a bound gadget $(T_\bougadl,\Tmodel_\bougadl)$.
% A subtree $M \in \Tmodel$ of $T$ (resp.\ $M \in \Tmodel_\bougadl$ of $T_\bougadl$) is said to be \emph{plugged into} $\selgad$ if $M$ intersects all subtrees in $\Tmodel_\selgadl$.
% The selection gadget $\selgad$ forces a solution of \textsc{Minimum Membership Dominating Set} on the instance $((T,\Tmodel),\ell)$ to include at least one of the models plugged into $\selgad$.

% Recall that exactly one of the models plugged into the selection gadget $\selgad$ is included in $\Tmodel^\ast$.
% Here, we require that each plugged-in model intersects all subtrees in $\Tmodel_\selgadl$.

\begin{restatable}{lemma}{lemselectgadget} \label{lem:select_gadget}
    % Let $\Tmodel^*$ be a solution to $((T,\Tmodel),\ell)$, and let $\Tmodel'$ be the set of models plugged into the selection gadget $\selgad = (T_\selgadl,\Tmodel_\selgadl)$.
    Let $((T,\Tmodel),\ell)$ be an instance of \textsc{Minimum Membership Dominating Set} and let $\selgad = (T_\selgadl,\Tmodel_\selgadl)$ be a selection gadget attached to some node in $T$.
    Suppose that at least one model is plugged into $\selgad$ and that every model in $\Tmodel \setminus \Tmodel_s$ that intersects some model in $\Tmodel_\selgadl$ is plugged into $\selgad$.
    If a solution $\Tmodel^\ast$ to $((T,\Tmodel),\ell)$ exists, then at least one of the models plugged into $\selgad$ is included in $\Tmodel^\ast$.
    Furthermore, $\Tmodel^\ast$ may be assumed to have no model in $\Tmodel_\selgadl$.
\end{restatable}

\begin{proof}
Let $\Tmodel'$ be the set of models plugged into the selection gadget $\selgad$.
For the sake of contradiction, assume that $\Tmodel^\ast \cap \Tmodel' = \emptyset$.
Since every model in $\Tmodel_\selgadl$ only intersects with the models in $\Tmodel'$, all the models in $\Tmodel_\selgadl$ must be contained in $\Tmodel^*$.
On the other hand, each model in $\Tmodel'$ intersects with all the models in $\Tmodel_\selgadl$.
This implies that for each model $M'$ in $\Tmodel'$, we have $\mem{T}{M'}{\Tmodel^\ast} \ge \ell+1$, which contradicts that $\Tmodel^\ast$ is a solution to $((T,\Tmodel),\ell)$.
Since at least one of $\Tmodel'$ is included in $\Tmodel^\ast$, it is clear that no model in $\Tmodel_\selgadl$ is required to belong to $\Tmodel^\ast$.
\end{proof}

% \begin{proof}
% Let $\Tmodel'$ be the set of models plugged into the selection gadget $\selgad$.
% For the sake of contradiction, assume that $\Tmodel^\ast \cap \Tmodel' = \emptyset$.
% Since every model in $\Tmodel_\selgadl$ only intersects with the models in $\Tmodel'$, all the models in $\Tmodel_\selgadl$ must be contained in $\Tmodel^*$.
% On the other hand, each model in $\Tmodel'$ intersects with all the models in $\Tmodel_\selgadl$.
% This implies that for each model $M'$ in $\Tmodel'$, we have $\mem{T}{M'}{\Tmodel^\ast} \ge \ell+1$, which contradicts that $\Tmodel^\ast$ is a solution to $((T,\Tmodel),\ell)$.
% Since at least one of $\Tmodel'$ is included in $\Tmodel^\ast$, it is clear that no model in $\Tmodel_\selgadl$ is required to belong to $\Tmodel^\ast$.
% \end{proof}

\subparagraph{Bound gadget.}
\begin{figure}[t]
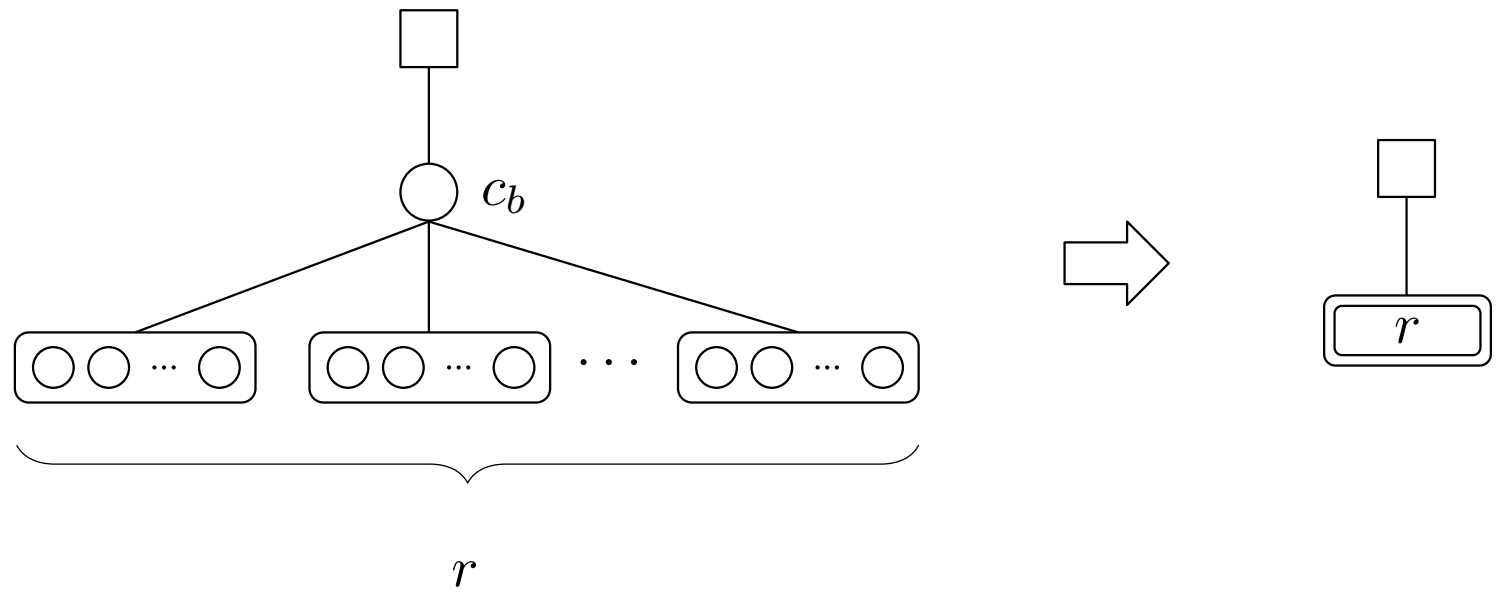

    \centering
    \resizebox{0.7\linewidth}{!}{\figboundgadget}
    \caption{The figure on the left illustrates the bound gadget $\bougad^{(r)}$, where the square denotes a node to which the gadget is attached. Note that the bound models in the gadget are omitted here. For simplicity, the bound gadget is represented as shown in the figure on the right.}
    \label{fig:bound_gadget}
\end{figure}

For a positive integer $r$, a bound gadget $\bougad^{(r)} = (T_\bougadl,\Tmodel_\bougadl)$ is defined as follows.
(See also \Cref{fig:bound_gadget}.)
The tree $T_\bougadl$ consists of a node $c_\bougadl$ and $r$ selection gadgets $\selgad_1,\ldots, \selgad_r$.
The center of each selection gadget is adjacent to $c_\bougadl$.
We call $c_\bougadl$ the \emph{center} of $T_\bougadl$.
The collection $\Tmodel_\bougadl$ contains the models of selection gadgets and $r$ \emph{bound models} $B_1,\ldots, B_r$.
For each $i \in [r]$, the bound model $B_i$ contains the center $c_\bougadl$ and is plugged into $\selgad_i$.
We say that a model $M$ is \emph{plugged into} $\bougad^{(r)}$ if $M$ contains only $c_\bougadl$ from $T_\bougadl$.

% We require that each plugged-in model contains the center of $\bougad^{(r)}$.

% A model $M$ for $(T, \Tmodel)$ is said to be \emph{plugged into} $\bougad^{(q)}$ if $M$ contains the center of $\bougad^{(q)}$.
% A bound gadget, which is plugged into a model $M$, aims to adjust the number of models that belong to a solution $M^\ast$ and intersect $M$.

\begin{restatable}{lemma}{lemauggadget} \label{lem:aug_gadget}
    Let $((T,\Tmodel),\ell)$ be an instance of \textsc{Minimum Membership Dominating Set} and let $\bougad^{(r)} = (T_\bougadl,\Tmodel_\bougadl)$ be a bound gadget attached to some node in $T$, where $r$ is a positive integer.
    Suppose that every model $M \in \Tmodel \setminus \Tmodel_b$ satisfies $M \cap V(T_\bougadl) \subseteq \{c_b\}$.
    If a solution $\Tmodel^\ast$ to $((T,\Tmodel),\ell)$ exists, then all bound models belong to $\Tmodel^\ast$, and hence $\mem{T}{M}{\Tmodel^\ast} \ge r$ for each model $M$ plugged into $\bougad^{(r)}$.
\end{restatable}

\begin{proof}
    Recall that for each $i \in [r]$, the bound model $B_i$ is plugged into the selection gadget $\selgad_i$.
    Thus, from \Cref{lem:select_gadget}, we have $B_i \in \Tmodel^\ast$ for each $i \in [r]$.
    Since the center of $\bougad^{(r)}$ is contained in $M$ and $B_i$ for each $i \in [r]$, we have $M \cap B_i \neq \emptyset$, which proves the lemma.
\end{proof}

% \begin{proof}
%     Recall that for each $i \in [r]$, the bound model $B_i$ is plugged into the selection gadget $\selgad_i$.
%     Thus, from \Cref{lem:select_gadget}, we have $B_i \in \Tmodel^\ast$ for each $i \in [r]$.
%     Since the center of $\bougad^{(r)}$ is contained in $M$ and $B_i$ for each $i \in [r]$, we have $M \cap B_i \neq \emptyset$, which proves the lemma.
% \end{proof}

\subparagraph{Grouped bound gadget.}
\begin{figure}[t]
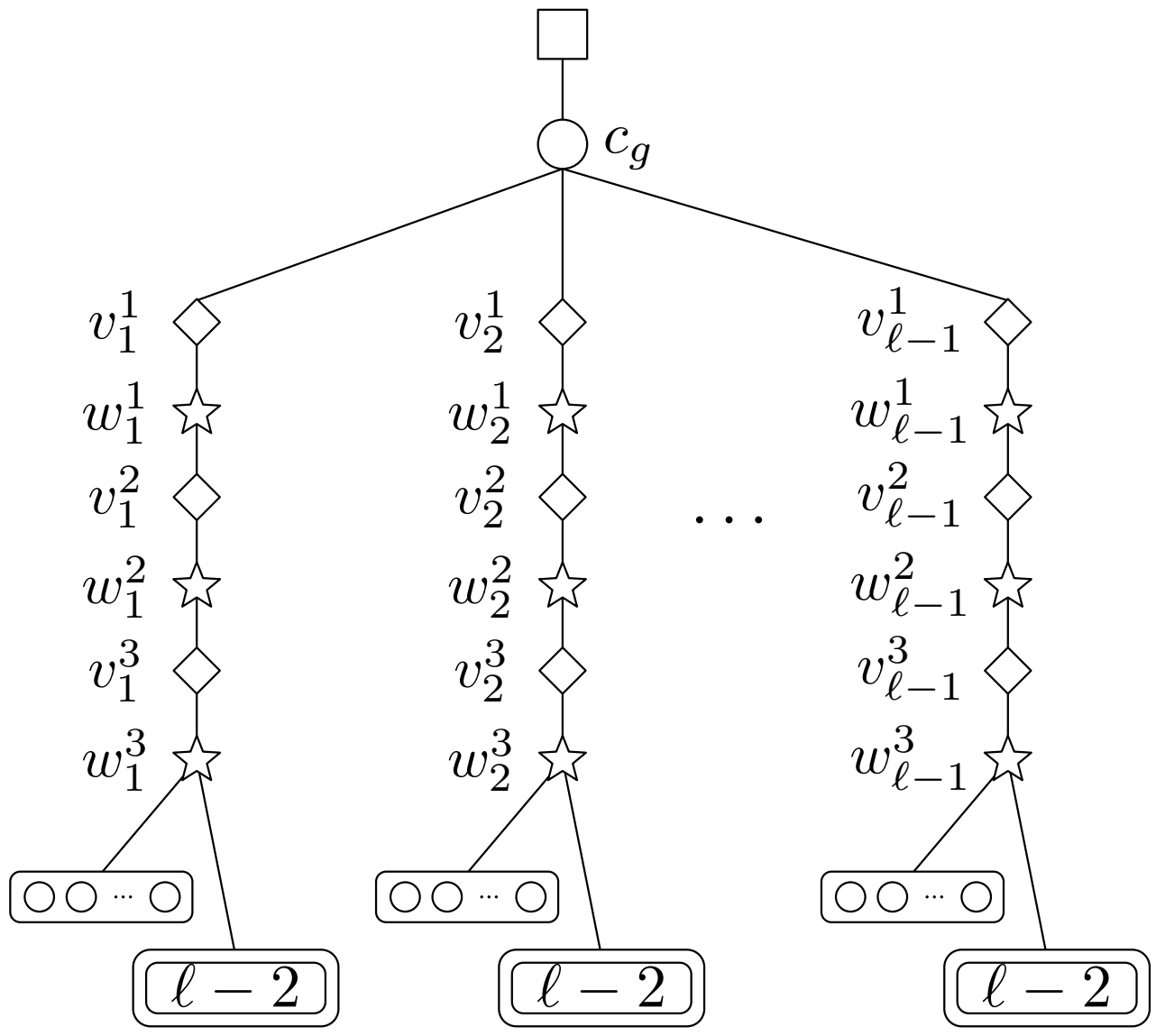

    \centering
    \resizebox{0.6\linewidth}{!}{\figgroupedboundgadget}
    \caption{The tree $T_\congadl$ of the grouped bound gadget $\congad$, where the square denotes a node to which the gadget is attached. Note that the models in the gadget are omitted here.}
    \label{fig:grouped_bound_gadget}
\end{figure}

% Let $r$ be a positive integer.
A grouped bound gadget is denoted by $\congad = (T_\congadl,\Tmodel_\congadl)$ and $T_\congadl$ has a node $c_\congadl$ called the \emph{center}.
(See also \Cref{fig:grouped_bound_gadget}.)
For $i \in [\ell-1]$, the path $P_{i}^\congadl$ is defined as $\langle c_\congadl, v_{i}^{1}, w_{i}^{1}, v_{i}^{2}, w_{i}^{2},\ldots, v_{i}^{n}, w_{i}^{n} \rangle$.
Furthermore, for each $i \in [\ell-1]$, the selection gadget $\selgad_i^{\congadl}$ and the bound gadget $\bougad_{i}^{\congadl,(\ell-2)}$ are introduced.
Each of $\selgad_i^{\congadl}$ and $\bougad_{i}^{\congadl,(\ell-2)}$ is attached to the node $w_{i}^{n}$.
This completes the construction of $T_\congadl$.

We then define $\Tmodel_\congadl$.
% The construction of $\Tmodel_\congadl$ is similar to that of $\Tmodel$.
First, all the models in the selection gadgets $\selgad_i^{\congadl}$ and the bound gadgets $\bougad_{i}^{\congadl,(\ell-2)}$ are included in $\Tmodel_\congadl$.
Next, for each node $t$ of $T_\congadl$ belonging to neither a selection gadget nor a bound gadget, add to $\Tmodel_\congadl$ a \emph{base model} consisting solely of $t$.
Then, for each pair of integers $i\in [\ell-1]$ and $j  \in [n]$, the \emph{group model} $F_{i}^j$ contains the subpath $\langle v_{i}^{j}, w_{i}^{j},\ldots, v_{i}^{n}, w_{i}^{n} \rangle$ and is plugged into the selection gadget $\selgad_i^{\congadl}$.
Finally, for each $i\in [\ell-1]$, add a \emph{restriction model} $\resmodel_{i}$ that contains the subpath $\langle v_{i}^{1}, \dots, w_{i}^{n} \rangle$ of $P_{i}^\congadl$ and is plugged into the bound gadget $\bougad_{i}^{\congadl,(\ell-2)}$.

% Suppose that $n$ models $M_1,\dots, M_n$ models are plugged into $\congad^{(r)}$.
% Here, for each $q \in [n]$, we say that the model $M_q$ is plugged into $\congad^{(r)}$ if $M_q$ consists of the subpath $\langle c_\congadl, v_{i}^{1}, w_{i}^{1} \dots, w_{i}^{q-1}, v_{i}^{q} \rangle$ of $P_{i}^\congadl$ for every $i \in [r]$.

Let $((T,\Tmodel),\ell)$ be an instance of \textsc{Minimum Membership Dominating Set}.
Whenever a grouped bound gadget $\congad = (T_\congadl,\Tmodel_\congadl)$ is used in $T$, it is assumed that exactly $n$ models $M_1,\dots, M_n$ in $\Tmodel \setminus \Tmodel_\congadl$ intersect with models in $\congad$.
More precisely, for each $q \in [n]$, we say that the model $M_q$ is \emph{plugged into} $\congad$ if $M_q$ contains the subpath $\langle c_\congadl, v_{i}^{1}, w_{i}^{1}, \dots, w_{i}^{q-1}, v_{i}^{q} \rangle$ and does not contain any nodes in the subpath $\langle w_{i}^{q}, \dots, v_{i}^{n}, w_{i}^{n} \rangle$ of $P_{i}^\congadl$ for every $i \in [\ell-1]$.

\begin{restatable}{lemma}{lemgroupedboundgadget} \label{lem:grouped_bound_gadget}
    \rev{Let $((T,\Tmodel),\ell)$ be an instance of \textsc{Minimum Membership Dominating Set} and let $\congad = (T_\congadl,\Tmodel_\congadl)$ be a grouped bound gadget attached to some node in $T$.
    Suppose that $\Tmodel^\ast$ is a solution to $((T,\Tmodel),\ell)$ containing at least one of $M_1,\ldots,M_n$.
    Then exactly one of these models, say $M_s$, belongs to $\mathcal M^\ast$.
    Furthermore, every model in $\{M_q: q\in [n], q \ge s \}$ intersects with exactly $\ell-1$ models belonging to $\Tmodel^\ast \cap \Tmodel_\congadl$.}
\end{restatable}

\begin{proof}
    Let $i\in [\ell-1]$. By \Cref{lem:select_gadget,lem:aug_gadget}, $\Tmodel^\ast$ contains at least one group model plugged into $\selgad_i^{\congadl}$ and the $\ell-2$ bound models in $\bougad_{i}^{\congadl,(\ell-2)}$.
    Since $\resmodel_{i}$ intersects with all these models, all of $M_1,\ldots,M_n$, and all base models on the subpath $\langle v_{i}^{1}, \dots, w_{i}^{n} \rangle$, we conclude from $\mem{T}{\resmodel_{i}}{\Tmodel^\ast} \le \ell$ that exactly one of $M_1,\ldots,M_n$, say $M_s$, and exactly one group model plugged into $\selgad_i^{\congadl}$ belong to $\Tmodel^\ast$, and that neither $\resmodel_{i}$ nor any of these base models belongs to $\Tmodel^\ast$.
    From the definition of $s$, for each $i\in [\ell-1]$, the node $v_{i}^s$ is contained in models in $\Tmodel^\ast \setminus \Tmodel_\congadl$, whereas the node $w_{i}^s$ is not.
    Furthermore, $\Tmodel^\ast$ has at least one model plugged into the selection gadget $\selgad_i^{\congadl}$ for each $i\in [\ell-1]$.
    Due to the restriction model $\resmodel_{i}$, the base model on the node $w_{i}^s$ is required to intersect with a group model $F_{i}^{s'} \in \Tmodel^\ast$ such that $s' \le s$. 
    Note that $F_{i}^{s'}$ and $M_q$ for every $q \ge s$ contain the node $v_{i}^s$.
    Moreover, the base model on the center $c_\congadl$ does not belong to $\Tmodel^\ast$.
    Otherwise, $M_s$ would intersect with the $\ell-1$ group models in $\Tmodel^\ast$, the base model on $c_\congadl$, and itself, and hence $\mem{T}{M_s}{\Tmodel^\ast} \ge \ell+1$, a contradiction.
    This completes the proof.
\end{proof}

% \begin{proof}
%     From the definition of $s$, for each $i\in [\ell-1]$, the node $v_{i}^s$ is contained in models in $\Tmodel^\ast \setminus \Tmodel_\congadl$, whereas the node $w_{i}^s$ is not.
%     Furthermore, $\Tmodel^\ast$ has at least one model plugged into the selection gadget $\selgad_i^{\congadl}$ for each $i\in [\ell-1]$.
%     Due to the restriction model $\resmodel_{i}$, the base model on the node $w_{i}^s$ is required to intersect with a group model $F_{i}^{s'} \in \Tmodel^\ast$ such that $s' \le s$. 
%     Note that $F_{i}^{s'}$ and $M_q$ for every $q \ge s$ contain the node $v_{i}^s$.
%     Thus, Claim (a) is shown.
%     In addition, if $s' > s$, then consider the set $\Tmodel' = \Tmodel^\ast \cup \{F_{i}^{s} \} \setminus \{F_{i}^{s'}\}$.
%     Observe that $\Tmodel'$ is also a solution to $((T,\Tmodel),\ell)$.
%     By applying this modification for every $i \in [\ell-1]$, Claim (b) is proved.
% \end{proof}

% \subsubsection{Correctness deferred from \Cref{app:correctness}}

\subsection{Construction of $T$} \label{subsec:construction_T}

We construct the tree $T$ of the tree representation $(T,\Tmodel)$ (see also \cref{fig:reduction}).\footnote{The symbols $\edgel$ and $\heightl$ stand for ``edge'' and ``height'', respectively. Intuitively, $P_{i,j}^\edgel$ controls the edge models, while $P_{i,j}^\heightl$ controls the height models. Furthermore, $\topl$ and $\bottoml$ denote ``top'' and ``bottom'', respectively, indicating whether they are located in the upper or lower part of $T$.}

First, add a node $c$, called the \emph{center} of $T$.
For each pair of two distinct integers $i, j \in [\numpart]$ with $i<j$, introduce the selection gadget $\selgad_{i,j}^{\edgel,\topl}$.
For each $i\in [\numpart]$, we also introduce the selection gadget $\selgad_{i}^{\heightl,\topl}$.
We attach all these selection gadgets to the center $c$.

Next, for each ordered pair of two distinct integers $i, j \in [\numpart]$, introduce two paths $P_{i,j}^\edgel$ and $P_{i,j}^\heightl$ defined as follows:
\begin{align*}
      P_{i,j}^\edgel & = \langle c, x_{i,j}^{\edgel,1}, y_{i,j}^{\edgel,1}, z_{i,j}^{\edgel,1}, x_{i,j}^{\edgel,2}, y_{i,j}^{\edgel,2}, z_{i,j}^{\edgel,2},\ldots, x_{i,j}^{\edgel,n}, y_{i,j}^{\edgel,n}, z_{i,j}^{\edgel,n} \rangle; \text{~and} \\
      P_{i,j}^\heightl & = \langle c, y_{i,j}^{\heightl,1}, x_{i,j}^{\heightl,1}, z_{i,j}^{\heightl,1}, y_{i,j}^{\heightl,2}, x_{i,j}^{\heightl,2}, z_{i,j}^{\heightl,2},\ldots, y_{i,j}^{\heightl,n}, x_{i,j}^{\heightl,n}, z_{i,j}^{\heightl,n} \rangle,
\end{align*}
where $x_{i,j}^{s,d}$, $y_{i,j}^{s,d}$, and $z_{i,j}^{s,d}$ are nodes of $T$ for each pair of $s \in \{\edgel, \heightl\}$ and $d \in [n]$ (see also \cref{fig:models}(a)).
Furthermore, we add the following three gadgets: the selection gadget $\selgad_{i,j}^{\edgel,\bottoml}$, the bound gadget $\bougad_{i,j}^{\edgel,\bottoml,(\ell-3)}$, and the grouped bound gadget $\congad_{i,j}^{\edgel}$.
Each of them is attached to the node $z_{i,j}^{\edgel,n}$.

Similarly, for each ordered pair of two distinct integers $i, j \in [\numpart]$, the selection gadget $\selgad_{i,j}^{\heightl,\bottoml}$, the bound gadget $\bougad_{i,j}^{\heightl,\bottoml,(\ell-3)}$, and the grouped bound gadget $\congad_{i,j}^{\heightl}$ are added, and each of them is attached to the node $z_{i,j}^{\heightl,n}$.
This completes the construction of $T$.

\subsection{Construction of $\Tmodel$} \label{subsec:construction_models}
% To complete the construction of the tree representation $(T, \Tmodel)$, this subsection defines $\Tmodel$.
First, $\Tmodel$ contains all the models associated with the gadgets.
The other models in $\Tmodel$ can be classified into five types: \emph{base models}, \emph{vertex models}, \emph{edge models}, \emph{height models}, and \emph{restriction models}, which are illustrated in \Cref{fig:models}. 

\subparagraph{Base models.}
For each node $t$ of $T$ other than $c$ that does not belong to any gadget, add to $\Tmodel$ a model consisting solely of $t$ (see \cref{fig:models}(a)).
We refer to these models as the \emph{base models}.

No base model can belong to any solution $\Tmodel^\ast \subseteq \Tmodel$ due to the other models defined below.
Therefore, the objective of \textsc{Minimum Membership Dominating Set} for $((T, \Tmodel), \ell)$ is to dominate all base models by other models in $\Tmodel^\ast$.

\subparagraph{Restriction models.}
For each ordered pair of two distinct integers $i, j \in [\numpart]$, add \emph{restriction models} $\resmodel_{i,j}^\edgel$ and $\resmodel_{i,j}^\heightl$ of $T$ into $\Tmodel$ (see \cref{fig:models}(b)).
The restriction model $\resmodel_{i,j}^\edgel$ contains the subpath $\langle x_{i,j}^{\edgel,1}, \dots, z_{i,j}^{\edgel,n} \rangle$ of $P_{i,j}^\edgel$ and is plugged into the bound gadget $\bougad_{i,j}^{\edgel,\bottoml,(\ell-3)}$.
Similarly, the restriction model $\resmodel_{i,j}^\heightl$ contains the subpath $\langle y_{i,j}^{\heightl,1}, \dots, z_{i,j}^{\heightl,n} \rangle$ of $P_{i,j}^\heightl$ and is plugged into the bound gadget $\bougad_{i,j}^{\heightl,\bottoml,(\ell-3)}$.

Due to the bound gadget, the restriction model $\resmodel_{i,j}^\edgel$ (resp.\ $\resmodel_{i,j}^\heightl$) intersects with at most three models in a solution $\Tmodel^\ast \subseteq \Tmodel$ on $P_{i,j}^\edgel$ (resp.\ $P_{i,j}^\heightl$).
This controls which models are included in $\Tmodel^\ast$.

\subparagraph{Vertex models.}
For each ordered pair of two distinct integers $i, j \in [\numpart]$ and each integer $q \in [n]$, add \emph{vertex models} $\vermodel_{i,j}^{\edgel,q}$ and $\vermodel_{i,j}^{\heightl,q}$ of $T$ to $\Tmodel$ (see \cref{fig:models}(c)).
The vertex model $\vermodel_{i,j}^{\edgel,q}$ contains the subpath $\langle z_{i,j}^{\edgel,q},\dots , z_{i,j}^{\edgel,n} \rangle$ of $P_{i,j}^\edgel$, is plugged into the selection gadget $\selgad_{i,j}^{\edgel,\bottoml}$, and is plugged into the grouped bound gadget $\congad_{i,j}^{\edgel}$.
Similarly, the vertex model $\vermodel_{i,j}^{\heightl,q}$ contains the subpath $\langle z_{i,j}^{\heightl,q},\dots , z_{i,j}^{\heightl,n} \rangle$ of $P_{i,j}^\heightl$, is plugged into the selection gadget $\selgad_{i,j}^{\heightl,\bottoml}$, and is plugged into the grouped bound gadget $\congad_{i,j}^{\heightl}$.

As the name suggests, for each $j\in [p]\setminus \{i\}$, the vertex models $\vermodel_{i,j}^{\edgel,q}$ and $\vermodel_{i,j}^{\heightl,q}$ correspond to the vertex $v_{i}^q$ of $G'$.
However, this is a many-to-one correspondence from the models 
$\{\vermodel_{i,j}^{\edgel,q},\vermodel_{i,j}^{\heightl,q}:j\in[p]\setminus\{i\}\}$ to $v_{i}^q$.
Consistency must be ensured by height models defined later.

\subparagraph{Edge models.}
Consider two integers $i, j \in [\numpart]$ with $i < j$.
For two integers $s,t \in [n]$, an \emph{edge model} $\edgemodel_{i,j}^{s,t}$ of $T$ is added to $\Tmodel$ if and only if $v_i^s v_j^t \in E_{i,j}$.
The edge model $\edgemodel_{i,j}^{s,t}$ is plugged into the selection gadget $\selgad_{i,j}^{\edgel,\topl}$ and contains the subpaths defined below:
the subpath $\langle c, x_{i,j}^{\edgel,1}, \dots, y_{i,j}^{\edgel,s} \rangle$ of $P_{i,j}^\edgel$; 
the subpath $\langle c,  y_{i,j}^{\heightl,1}, \dots, y_{i,j}^{\heightl,s} \rangle$ of $P_{i,j}^\heightl$;
the subpath $\langle c, x_{j,i}^{\edgel,1}, \dots, y_{j,i}^{\edgel,t} \rangle$ of $P_{j,i}^\edgel$; and
the subpath $\langle c, y_{j,i}^{\heightl,1}, \dots, y_{j,i}^{\heightl,t} \rangle$ of $P_{j,i}^\heightl$ (see \cref{fig:models}(d)).

% \begin{itemize}
%   \item the subpath $\langle c, x_{i,j}^{\edgel,1}, \dots, y_{i,j}^{\edgel,s} \rangle$ of $P_{i,j}^\edgel$; 
%   \item the subpath $\langle c,  y_{i,j}^{\heightl,1}, \dots, y_{i,j}^{\heightl,s} \rangle$ of $P_{i,j}^\heightl$;
%   \item the subpath $\langle c, x_{j,i}^{\edgel,1}, \dots, y_{j,i}^{\edgel,t} \rangle$ of $P_{j,i}^\edgel$; and
%   \item the subpath $\langle c, y_{j,i}^{\heightl,1}, \dots, y_{j,i}^{\heightl,t} \rangle$ of $P_{j,i}^\heightl$.
% \end{itemize}

The edge models represent the edges of $G'$.
We will show that $G'$ has a clique $C$ satisfying the conditions of \textsc{Multicolored Clique} if and only if edge models corresponding to edges in $C$ are included in a solution $\Tmodel^\ast \subseteq \Tmodel$ of \textsc{Minimum Membership Dominating Set}.

\subparagraph{Height models.}
For two integers $i \in [\numpart]$ and $s \in [n]$, add a \emph{height model} $\himodel_{i}^{s}$ of $T$ into $\Tmodel$.
The height model $\himodel_{i}^{s}$ contains the subpath $\langle c, x_{i,j}^{\edgel,1}, \dots, x_{i,j}^{\edgel,s} \rangle$ of $P_{i,j}^\edgel$ and the subpath $\langle c,  y_{i,j}^{\heightl,1}, \dots, x_{i,j}^{\heightl,s} \rangle$ of $P_{i,j}^\heightl$ for each $j \in [\numpart] \setminus \{i\}$ (see \cref{fig:models}(e)).
Moreover, the height model $\himodel_{i}^{s}$ is plugged into the selection gadget $\selgad_{i}^{\heightl,\topl}$.

This model ensures that for two distinct integers $j_1, j_2 \in [\numpart] \setminus \{ i \}$, if $\vermodel_{i,j_1}^{\edgel,q_1}, \vermodel_{i,j_2}^{\edgel,q_2} \in \Tmodel^\ast$, then we have $q_1 = q_2$.

\subsection{The leafage of $G$} \label{subsec:num_leafage}
Let $G$ be a chordal graph corresponding to the tree representation $(T,\Tmodel)$.
We count the number of leaves in $T$ and bound the leafage of $G$.

From the construction of $G$, all leaves in $T$ are contained in selection gadgets, bound gadgets, and grouped bound gadgets.
Moreover, all leaves of the grouped bound gadgets are derived from selection gadgets and bound gadgets, and all leaves of the bound gadgets are derived from selection gadgets.
Therefore, we count the number of selection gadgets in $T$.

For each $i\in [\numpart]$, there is a selection gadget $\selgad_{i}^{\heightl,\topl}$, and for each pair of integers $i, j \in [\numpart]$ with $i<j$, there is a selection gadget $\selgad_{i,j}^{\edgel,\topl}$.
Let $(i,j)$ be an ordered pair of two distinct integers $i, j \in [\numpart]$.
There are two selection gadgets $\selgad_{i,j}^{\edgel,\bottoml}$ and $\selgad_{i,j}^{\heightl,\bottoml}$.
The tree $T$ contains two bound gadgets $\bougad_{i,j}^{\edgel,\bottoml,(\ell-3)}$ and $\bougad_{i,j}^{\heightl,\bottoml,(\ell-3)}$, which yield $2\ell -6$ selection gadgets in total.
Furthermore, the tree has two grouped bound gadgets $\congad_{i,j}^{\edgel}$ and $\congad_{i,j}^{\heightl}$.
Each grouped bound gadget has $\ell-1$ selection gadgets and $\ell-1$ bound gadgets.
Since each bound gadget has $\ell-2$ selection gadgets, the number of selection gadgets in $\congad$ is $(\ell-1) + (\ell-1)(\ell-2) = (\ell-1)^2$.

Thus, the total number of selection gadgets in $T$ is as follows:
\[
  \numpart + \numpart(\numpart-1)/2 + \numpart(\numpart-1)(2+ (2\ell-6) + 2(\ell-1)^2).
\]
Since each selection gadget has $\ell+1$ leaves and $\ell$ is bounded by a function of $\numpart$, the leafage of $G$ is bounded by a function of $\numpart$.

\subsection{Correctness of our reduction} \label{subsec:correctness}
It is clear that the tree representation $(T,\Tmodel)$ and its corresponding graph $G$ can be constructed in polynomial time.
We show that $(G', \numpart, (V_1, V_2, \ldots, V_\numpart))$ is a yes-instance of \textsc{Multicolored Clique} if and only if $((T,\Tmodel), \ell = 3\numpart^2)$ is a yes-instance of \textsc{Minimum Membership Dominating Set}.

We first show the sufficiency.
Suppose that $G'$ has a clique $C = \{v_1^{q_1}, v_2^{q_2}, \dots, v_\numpart^{q_\numpart} \}$, where $q_1, q_2,\dots ,q_\numpart \in [n]$.
Let $\Tmodel_1 \subseteq \Tmodel$ be a set of models defined as follows:
\[
  \Tmodel_1 = \{\vermodel_{i,j}^{\edgel,q_i}, \vermodel_{i,j}^{\heightl,q_i} : i,j \in [\numpart], i \neq j  \}  \cup \{ \edgemodel_{i,j}^{q_i,q_j} : i,j \in [\numpart], i < j \} \cup \{ \himodel_{i}^{q_i} : i \in [\numpart] \}.
\]

We also define $\Tmodel_2 \subseteq \Tmodel$ as the set of ``suitable'' models in the gadgets that enable them to operate as intended.
Formally define $\Tmodel_2$ as follows.
Recall that there exists the grouped bound gadget $\congad_{i,j}^{\edgel}$ for each pair $i, j \in [p]$ with $i \neq j$.
For each $r \in [\ell-1]$, we add the group model $F_{r}^{q_i}$ in $\congad_{i,j}^{\edgel}$ to $\Tmodel_2$.
The same additions are performed for the grouped bound gadget $\congad_{i,j}^{\heightl}$.
Furthermore, we add the bound models in all the bound gadgets to $\Tmodel_2$.
The construction of $\Tmodel_2$ is completed.
Finally, let $\Tmodel^\ast = \Tmodel_1 \cup \Tmodel_2$.

To show that $\Tmodel^\ast$ is a solution to $((T,\Tmodel), \ell)$, it suffices to verify the following two conditions:
    \begin{enumerate}[(1)]
        \item for each $M \in \Tmodel$, there exists a model $M^\ast \in \Tmodel^\ast$ such that $M \cap M^\ast \neq \emptyset$; and  
        \item $\mem{T}{M}{\Tmodel^\ast} \le \ell $ for every model $M \in  \Tmodel$. 
    \end{enumerate}
% However, we have not yet defined the models in the gadgets, and therefore cannot verify whether all models satisfy the above conditions.
% In the following, assuming that the models in the gadgets satisfy these conditions, we demonstrate that the conditions are satisfied for base models, vertex models, edge models, height models, and restriction models.

We begin by arguing about the models in the gadgets.

\subparagraph{Selection gadget.}
There are four types of selection gadgets that are not part of bound gadgets and grouped bound gadgets: $\selgad_{i,j}^{\edgel,\bottoml}$, $\selgad_{i,j}^{\heightl,\bottoml}$, $\selgad_{i,j}^{\edgel,\topl}$, and $\selgad_{i}^{\heightl,\topl}$, where $i, j \in [p]$.
% We show that each selection model $M$ in these selection gadgets satisfies Conditions~(1) and~(2).
We only explain that each selection model $M$ in $\selgad_{i,j}^{\edgel,\bottoml}$ satisfies Conditions~(1) and~(2) because similar arguments can be made for the other selection gadgets.

Consider the selection gadget $\selgad_{i,j}^{\edgel,\bottoml}$, where $i, j \in [p]$.
As described in \Cref{subsec:construction_models}, the vertex models $\vermodel_{i,j}^{\edgel,1}, \dots, \vermodel_{i,j}^{\edgel,n}$ are plugged into $\selgad_{i,j}^{\edgel,\bottoml}$ (that is, contain all selection models in $\selgad_{i,j}^{\edgel,\bottoml}$), whereas the other models are not.
Since $\vermodel_{i,j}^{\edgel,q_i} \in \Tmodel^\ast$, Condition~(1) is satisfied.
Moreover, no selection model and no vertex model other than $\vermodel_{i,j}^{\edgel,q_i}$ belongs to $\Tmodel^\ast$.
Thus, we have $\mem{T}{M}{\Tmodel^\ast} = 1 \le \ell$, and Condition~(2) is satisfied.

% Similarly, the vertex model $\vermodel_{i,j}^{\heightl,q_i} \in \Tmodel^\ast$, the edge model $\edgemodel_{i,j}^{q_i,q_j} \in \Tmodel^\ast$, and the height model $\himodel_{i}^{q_i} \in \Tmodel^\ast$ are plugged into $\selgad_{i,j}^{\heightl,\bottoml}$, $\selgad_{i,j}^{\edgel,\topl}$, and $\selgad_{i}^{\heightl,\topl}$, respectively.
% Similarly, the vertex model $\vermodel_{i,j}^{\heightl,q_i}$, the edge model $\edgemodel_{i,j}^{q_i,q_j}$, and the height model $\himodel_{i}^{q_i}$ are the only models in $\Tmodel^\ast$ that are plugged into $\selgad_{i,j}^{\heightl,\bottoml}$, $\selgad_{i,j}^{\edgel,\topl}$, and $\selgad_{i}^{\heightl,\topl}$, respectively.
% Following the same arguments as above, we can show that Conditions~(1) and~(2) are satisfied.

\subparagraph{Bound gadget.}
There are two types of bound gadgets that are not part of grouped bound gadgets: $\bougad_{i,j}^{\edgel,\bottoml,(\ell-3)}$ and $\bougad_{i,j}^{\heightl,\bottoml,(\ell-3)}$, where $i, j \in [p]$.
We only explain that each model $M$ in $\bougad_{i,j}^{\edgel,\bottoml,(\ell-3)}$ satisfies Conditions~(1) and~(2) because similar arguments can be made for the other bound gadgets.

Recall that $\bougad_{i,j}^{\edgel,\bottoml,(\ell-3)}$ has the bound models $B_1, \dots, B_{\ell-3}$ and the selection models in the selection gadgets $\selgad_1,\ldots, \selgad_{\ell-3}$.
For each $r \in [\ell-3]$, the bound model $B_r$ belongs to $\Tmodel_2 \subseteq \Tmodel^\ast$ and is plugged into $\selgad_r$.
Thus, Condition~(1) is satisfied.

Consider Condition~(2).
Let $r \in [\ell-3]$.
For each selection model $M$ in the selection gadget $\selgad_r$, we have $\mem{T}{M}{\Tmodel^\ast} = 1 \le \ell$, analogous to the explanation above.
Recall that $B_r$ intersects $B_s$ for each $s \in [\ell-3] \setminus \{r\}$, all the selection models in $\selgad_r$, and the restriction model $\resmodel_{i,j}^\edgel$. 
All bound models belong to $\Tmodel_2$, whereas no selection model does.
Furthermore, $\resmodel_{i,j}^\edgel \notin \Tmodel^\ast$. 
Therefore, for each $r \in [\ell-3]$, we have $\mem{T}{B_r}{\Tmodel^\ast} = \ell-3 \le \ell$, and hence Condition~(2) is satisfied.

% Similarly, we can prove that the models in  $\bougad_{i,j}^{\edgel,\bottoml,(\ell-3)}$, 
% $\bougad_{i,j}^{\heightl,\bottoml,(\ell-1)}$, and $\bougad_{i,j}^{\heightl,\bottoml,(\ell-3)}$ also satisfy Conditions~(1) and~(2).

\subparagraph{Grouped bound gadget.}
Consider the grouped bound gadget $\congad_{i,j}^{\edgel}$, which is attached to the node $z_{i,j}^{\edgel,n}$ of the path $P_{i,j}^\edgel$, where $i, j \in [p]$.
The above argument applies to the models belonging to the selection gadgets and the bound gadgets in $\congad_{i,j}^{\edgel}$.
Recall that the other models in $\congad_{i,j}^{\edgel}$ are classified into the following three types: base models, group models, and restriction models.
Let $r \in [\ell-1]$.

% Next, for each node $t$ of $T_\congadl$ not belonging to any selection gadget and bound gadget, add to $\Tmodel_\congadl$ a \emph{base model} consisting solely of $t$.
% Then, for each pair of integers $i\in [r]$ and $j  \in [n]$, the \emph{group model} $F_{i}^j$ contains the subpath $\langle v_{i}^{j}, w_{i}^{j},\ldots, v_{i}^{n}, w_{i}^{n} \rangle$ and is plugged into the selection gadget $\selgad_i^{\congadl}$.
% Finally, add a \emph{restriction model} $\resmodel_{i}$ that contains the subpath $\langle v_{i}^{1}, \dots, w_{i}^{n} \rangle$ of $P_{i}^\congadl$ and is plugged into the bound gadget $\bougad_{i}^{\congadl,(\ell-2)}$.

Recall that the path $P_{r}^\congadl$ is defined as $\langle c_\congadl, v_{r}^{1}, w_{r}^{1}, v_{r}^{2}, w_{r}^{2},\ldots, v_{r}^{n}, w_{r}^{n} \rangle$.
From the definitions of base models, group models, and restriction models, each of these models contains at least one node from $P_{r}^\congadl$.
Moreover, each node belonging to $P_{r}^\congadl$ intersects with $\vermodel_{i,j}^{\edgel,q_i} \in \Tmodel^\ast$ or $F_{r}^{q_i} \in \Tmodel^\ast$.
Thus, Condition~(1) is satisfied.

Consider Condition~(2).
Each of the base models and group models on $P_{r}^\congadl$ intersects with only two models in $\Tmodel^\ast$: one is the vertex model $\vermodel_{i,j}^{\edgel,q_i}$ and the other is the group model $F_{r}^{q_i}$.
The restriction model $\resmodel_{r}$ additionally intersects $\ell-2$ bound models, which belong to $\Tmodel^\ast$, in the bound gadget $\bougad_{r}^{\congadl,(\ell-2)}$.
Therefore, Condition~(2) is satisfied.

The same arguments apply to the grouped bound gadget $\congad_{i,j}^{\heightl}$.

\vspace{\baselineskip}

We also show that the models outside the gadgets satisfy Conditions~(1) and~(2).

\subparagraph{Base models.}
For each ordered pair of two distinct integers $i, j \in [\numpart]$, each base model on $P_{i,j}^\edgel$ intersects with at least one and at most three models that belong to $\Tmodel^\ast$: $\vermodel_{i,j}^{\edgel,q_i}$, $\edgemodel_{i,j}^{q_i,q_j}$, and $\himodel_{i}^{q_i}$.
Since $\ell > 3$ from the fact that $\ell = 3p^2$ and $p\ge 2$, Conditions~(1) and~(2) are satisfied.
The same arguments are applied to the base models on $P_{i,j}^\heightl$.

\subparagraph{Restriction models.}
Consider the restriction model $\resmodel_{i,j}^\edgel$, where $i$ and $j$ are two distinct integers in $[\numpart]$. 
Recall that $\resmodel_{i,j}^\edgel$ is plugged into the bound gadget $\bougad_{i,j}^{\edgel,\bottoml,(\ell-3)}$.
Thus, there are $\ell-3$ models in $\bougad_{i,j}^{\edgel,\bottoml,(\ell-3)}$ that belong to $\Tmodel^\ast$ and intersect with $\resmodel_{i,j}^\edgel$.
Moreover, $\resmodel_{i,j}^\edgel$ intersects with three additional models in $\Tmodel^\ast$: $\vermodel_{i,j}^{\edgel,q_i}$, $\edgemodel_{i,j}^{q_i,q_j}$, and $\himodel_{i}^{q_i}$.
Therefore, Conditions~(1) and~(2) are satisfied.
The proof for $\resmodel_{i,j}^\heightl$ follows in the same way.

\subparagraph{Vertex models.}
Let $i$ and $j$ be two distinct integers in $[\numpart]$, and let $q \in [n]$.
The vertex model $\vermodel_{i,j}^{\edgel,q}$ contains the subpath $\langle z_{i,j}^{\edgel,q},\dots , z_{i,j}^{\edgel,n} \rangle$ of $P_{i,j}^\edgel$, and hence intersects with $\vermodel_{i,j}^{\edgel,q_i} \in \Tmodel^\ast$ at the node $z_{i,j}^{\edgel,n}$.
Thus, Condition~(1) is satisfied.

Consider next Condition~(2).
The vertex model $\vermodel_{i,j}^{\edgel,q}$ is plugged into the selection gadget $\selgad_{i,j}^{\edgel,\bottoml}$ and the grouped bound gadget $\congad_{i,j}^{\edgel}$.
No model in $\selgad_{i,j}^{\edgel,\bottoml}$ belongs to $\Tmodel^\ast$ by the definition of $\Tmodel^\ast$.
Among the models in $\Tmodel_2$, the vertex model $\vermodel_{i,j}^{\edgel,q}$ can intersect only with the group models $F_{r}^{q_i}$ in $\congad_{i,j}^{\edgel}$, where $r \in [\ell-1]$.
If $q \ge q_i$, then $\vermodel_{i,j}^{\edgel,q}$ contains the node $v_{r}^{q_i}$ for every $r \in [\ell-1]$, and hence intersects with all these $\ell-1$ group models.
Since $\vermodel_{i,j}^{\edgel,q_i}$ is the unique model in $\Tmodel_1$ that intersects with $\vermodel_{i,j}^{\edgel,q}$, we have $\mem{T}{\vermodel_{i,j}^{\edgel,q}}{\Tmodel^\ast} = \ell$.
If $q < q_i$, then $\vermodel_{i,j}^{\edgel,q}$ intersects with none of the group models $F_{r}^{q_i}$.
Since $\vermodel_{i,j}^{\edgel,q}$ can intersect only with $\vermodel_{i,j}^{\edgel,q_i}$, $\edgemodel_{i,j}^{q_i,q_j}$, and $\himodel_{i}^{q_i}$ among the models in $\Tmodel_1$, we have $\mem{T}{\vermodel_{i,j}^{\edgel,q}}{\Tmodel^\ast} \le 3 \le \ell$.
Thus, Condition~(2) is satisfied in both cases.
The proof for $\vermodel_{i,j}^{\heightl,q}$ is analogous.

\subparagraph{Edge models.}
Since every edge model contains the center $c$ and there is at least one edge model in $\Tmodel^\ast$, Condition~(1) is satisfied.

Let $i,j \in [\numpart]$ with $i < j$, and let $s, t \in [n]$.
Recall that the edge model $\edgemodel_{i,j}^{s,t}$ is plugged into the selection gadget $\selgad_{i,j}^{\edgel,\topl}$.
By the definition of $\Tmodel^\ast$, no selection model in $\selgad_{i,j}^{\edgel,\topl}$ belongs to $\Tmodel^\ast$.
% there is no model in gadgets that belongs to $\Tmodel^\ast$ and intersects with $\edgemodel_{i,j}^{s,t}$.
There are at most $\numpart(\numpart-1)/2$ edge models and exactly $\numpart$ height models that intersect with $\edgemodel_{i,j}^{s,t}$ because all of them contain the center $c$.
Furthermore, on each of the four paths $P_{i,j}^\edgel$, $P_{i,j}^\heightl$, $P_{j,i}^\edgel$, and $P_{j,i}^\heightl$, at most one vertex model intersects with $\edgemodel_{i,j}^{s,t}$.
Consequently, we have
\[
  \mem{T}{\edgemodel_{i,j}^{s,t}}{\Tmodel^\ast} \le \numpart(\numpart-1)/2 + \numpart + 4 = (\numpart^2+\numpart+8)/2 \le 3\numpart^2 = \ell,
\]
and hence Condition~(2) is satisfied.

% For two integers $s,t \in [n]$, an \emph{edge model} $\edgemodel_{i,j}^{s,t}$ of $T$ is added to $\Tmodel$ if and only if $v_i^s v_j^t \in E_{i,j}$.
% The edge model $\edgemodel_{i,j}^{s,t}$ is plugged into the selection gadget $\selgad_{i,j}^{\edgel,\topl}$ and contains the subpaths defined below:
% \begin{itemize}
%   \item the subpath $\langle c, x_{i,j}^{\edgel,1}, \dots, y_{i,j}^{\edgel,s} \rangle$ of $P_{i,j}^\edgel$, 
%   \item the subpath $\langle c,  y_{i,j}^{\heightl,1}, \dots, y_{i,j}^{\heightl,s} \rangle$ of $P_{i,j}^\heightl$,
%   \item the subpath $\langle c, x_{j,i}^{\edgel,1}, \dots, y_{j,i}^{\edgel,t} \rangle$ of $P_{j,i}^\edgel$, and
%   \item the subpath $\langle c, y_{j,i}^{\heightl,1}, \dots, y_{j,i}^{\heightl,t} \rangle$ of $P_{j,i}^\heightl$.
% \end{itemize}

\subparagraph{Height models.}
This case is similar to the case of edge models.
Observe that Condition~(1) is satisfied.
Let $i \in [\numpart]$ and let $s \in [n]$.
Recall that the height model $\himodel_{i}^{s}$ is plugged into the selection gadget $\selgad_{i}^{\heightl,\topl}$.
By the definition of $\Tmodel^\ast$, no selection model in $\selgad_{i}^{\heightl,\topl}$ belongs to $\Tmodel^\ast$.
% we may assume that there is no model in gadgets that belongs to $\Tmodel^\ast$ and intersects with $\himodel_{i}^{s}$.
There are exactly $\numpart(\numpart-1)/2$ edge models and at most $\numpart$ height models that intersect with $\himodel_{i}^{s}$ because all of them contain the center $c$.
Furthermore, on each of the $2(p-1)$ paths $P_{i,1}^\edgel, P_{i,1}^\heightl, P_{i,2}^\edgel, P_{i,2}^\heightl, \ldots, P_{i,\numpart}^\edgel$, and $P_{i,\numpart}^\heightl$, at most one vertex model intersects with $\himodel_{i}^{s}$.
Consequently, we have
\[
  \mem{T}{\himodel_{i}^{s}}{\Tmodel^\ast} \le \numpart(\numpart-1)/2 + \numpart + 2(\numpart-1) = (\numpart^2+5\numpart-4)/2 \le 3\numpart^2 = \ell,
\]
and hence Condition~(2) is satisfied.

\vspace{\baselineskip}

In summary, Conditions~(1) and~(2) are satisfied for all models.

% Recall that $\Tmodel^\ast = \Tmodel_1 \cup \Tmodel_2$.
% In addition, it suffices to verify the following two conditions:
% \begin{enumerate}[(1)]
%     \item for each $M \in \Tmodel$, there exists a model $M^\ast \in \Tmodel^\ast$ such that $M \cap M^\ast \neq \emptyset$; and  
%     \item $\mem{T}{M}{\Tmodel^\ast} \le \ell $ for every model $M \in  \Tmodel$. 
% \end{enumerate}
% We now show that these conditions are satisfied for models in the gadgets.

We next show the necessity.
Suppose that there exists a set $\Tmodel^\ast \subseteq \Tmodel$ such that Conditions~(1) and~(2) are satisfied.
% Let $\Tmodel_1$ be a subset of $\Tmodel^\ast$ consisting of vertex models, edge models, and height models.
Recall that every vertex model on $P_{i,j}^\edgel$ is plugged into the selection gadget $\selgad_{i,j}^{\edgel,\bottoml}$, where $i$ and $j$ are two distinct integers in $[\numpart]$.
From \Cref{lem:select_gadget}, at least one vertex model on $P_{i,j}^\edgel$ is included in $\Tmodel^\ast$.
Similarly, for each pair of integers $i,j \in [\numpart]$ with $i < j$, at least one edge model is included in $\Tmodel^\ast$, and for each $i \in [\numpart]$, at least one height model is included in $\Tmodel^\ast$.
This implies that for each pair of distinct integers $i, j \in [\numpart]$, there exists at least one vertex model, at least one edge model, and at least one height model that belong to $\Tmodel^\ast$ and intersect the restriction model $\resmodel_{i,j}^\edgel$.
In fact, since the restriction model $\resmodel_{i,j}^\edgel$ intersects with $\ell-3$ bound models in the bound gadget $\bougad_{i,j}^{\edgel,\bottoml,(\ell-3)}$, these three models are uniquely determined for each pair of distinct integers $i, j \in [\numpart]$.
In other words, $\Tmodel^\ast$ contains neither a base model nor a restriction model.
The same arguments hold for $\resmodel_{i,j}^\heightl$.

The above discussion indicates that for each pair of distinct integers $i,j \in [\numpart]$, there is a unique integer $q\in [n]$ such that $\vermodel_{i,j}^{\heightl,q} \in \Tmodel^\ast$.
Here, we give the following two lemmas.

\begin{lemma} \label{applem:heightj}
    Let $i \in [\numpart]$ and let $j, j' \in [\numpart] \setminus \{i\}$.
    For two integers $q, q' \in [n]$, if $\vermodel_{i,j}^{\heightl,q} \in \Tmodel^\ast$ and $\vermodel_{i,j'}^{\heightl,q'} \in \Tmodel^\ast$, then $q = q'$.
\end{lemma}

\begin{proof}   
    Suppose that $\vermodel_{i,j}^{\heightl,q} \in \Tmodel^\ast$.
    Recall that $\vermodel_{i,j}^{\heightl,q}$ is the unique vertex model on $P_{i,j}^\heightl$ that belongs to $\Tmodel^\ast$.
    Since $\vermodel_{i,j}^{\heightl,q}$ does not contain $x_{i,j}^{\heightl,q}$ and $\Tmodel^\ast$ contains neither any base model nor any restriction model, there is an edge model or a height model $M \in \Tmodel^\ast$ that contains $x_{i,j}^{\heightl,q}$.
    
    For the sake of contradiction, assume that $M$ is an edge model.
    By the definition of edge models, $M$ also contains $z_{i,j}^{\heightl,q}$.
    Furthermore, recall that $\vermodel_{i,j}^{\heightl,q}$ contains $z_{i,j}^{\heightl,q}$.
    Hence, $M$ intersects with $\vermodel_{i,j}^{\heightl,q}$.
    However, this leads to a contradiction because we have $\mem{T}{\vermodel_{i,j}^{\heightl,q}}{\Tmodel^\ast} > \ell$ due to the $(\ell-1)$ models in the grouped bound gadget $\congad_{i,j}^{\heightl}$, the model $M$, and itself.
    Thus, $M$ is a height model, and hence $M = \himodel_{i}^{r}$ for some $r \ge q$ because $M$ contains $x_{i,j}^{\heightl,q}$.
    If $r > q$, then $\himodel_{i}^{r}$ also contains $z_{i,j}^{\heightl,q}$, which leads to the same contradiction.
    Thus, we have $M = \himodel_{i}^{q}$ and $\himodel_{i}^{q} \in \Tmodel^\ast$.

    Similarly, $\himodel_{i}^{q'} \in \Tmodel^\ast$ holds.
    Since exactly one height model intersecting $\resmodel_{i,j}^\edgel$ belongs to $\Tmodel^\ast$ as shown above, the height model $\himodel_{i}^{q}$ must be identical to the height model $\himodel_{i}^{q'}$, leading to $q = q'$.
\end{proof}

\begin{lemma} \label{applem:heighteh}
    Let $i, j \in [\numpart]$ with $i\neq j$.
    For two integers $q, q' \in [n]$, if $\vermodel_{i,j}^{\heightl,q} \in \Tmodel^\ast$ and $\vermodel_{i,j}^{\edgel,q'} \in \Tmodel^\ast$, then $q = q'$.
\end{lemma}

\begin{proof}
    Assume that $i < j$.
    The case $i > j$ can be proved analogously.
    As in the proof of \Cref{applem:heightj}, $\vermodel_{i,j}^{\heightl,q} \in \Tmodel^\ast$ implies that $\himodel_{i}^{q} \in \Tmodel^\ast$.
    By \Cref{lem:grouped_bound_gadget}, each of $\vermodel_{i,j}^{\heightl,q}$ and $\vermodel_{i,j}^{\edgel,q'}$ intersects with $\ell-1$ models in its grouped bound gadget and with itself, and hence with no other model in $\Tmodel^\ast$.
    Since $\vermodel_{i,j}^{\edgel,q'}$ cannot intersect with $\himodel_{i}^{q}$, we have $q \le q'$.
    Moreover, $\vermodel_{i,j}^{\heightl,q}$ cannot intersect with the edge model $\edgemodel_{i,j}^{s,t} \in \Tmodel^\ast$, where $s,t \in [n]$.
    Thus, we have $s \le q$.
    This implies that $\edgemodel_{i,j}^{s,t}$ does not contain $z_{i,j}^{\edgel,q}$.
    Since $\Tmodel^\ast$ contains neither any base model nor any restriction model, and $\himodel_{i}^{q}$ and $\edgemodel_{i,j}^{s,t}$ are the unique height and edge models in $\Tmodel^\ast$ that intersect $\resmodel_{i,j}^\edgel$, the base model on $z_{i,j}^{\edgel,q}$ must be dominated by the vertex model $\vermodel_{i,j}^{\edgel,q'} \in \Tmodel^\ast$.
    This leads to $q \ge q'$.
    Since $q \le q'$ as shown above, we conclude that $q = q'$.
\end{proof}

From \Cref{applem:heightj,applem:heighteh}, for each $i \in [\numpart]$, there exists an integer $q_i \in [n]$ such that $\vermodel_{i,j}^{\edgel,q_i}, \vermodel_{i,j}^{\heightl,q_i} \in \Tmodel^\ast$, where $j \in [\numpart] \setminus \{i\}$.
Let $C = \{v_i^{q_i} :  i \in [\numpart] \}$.
We show that $C$ forms a clique of $G'$.

% As described in the proof of \Cref{applem:heightj}, we have $\himodel_{i}^{q_i} \in \Tmodel^\ast$ for each $i \in [\numpart]$. 
% Thus, due to the grouped bound gadget $\congad_{i,j}^{\heightl,\bottoml,(\ell-1)}$, there exists an integer $q' \in [n]$ such that the edge model $\edgemodel_{i,j}^{q_i,q'}\in \Tmodel^\ast$ intersects the base model on $y_{i,j}^{\edgel,q_i}$ (assuming $i < j$).

% Consider the path $P_{j,i}^\edgel$.
% As with the above discussion, the base model on $y_{j,i}^{\edgel,q_j}$ must be dominated by the edge model $\edgemodel_{i,j}^{q_i,q'}\in \Tmodel^\ast$, which implies that $q' = q_j$.
% Recall that $\edgemodel_{i,j}^{q_i,q_j}\in \Tmodel$ if and only if $v_i^{q_i} v_j^{q_j} \in E_{i,j}$.
% Therefore, $C$ is a clique of $G'$.
% Since $|C \cap V_i| =1$ for each $i\in [\numpart]$, $(G', \numpart, (V_1, V_2, \ldots, V_\numpart))$ is a yes-instance of \textsc{Multicolored Clique}.

As described in the proof of \Cref{applem:heightj}, we have $\himodel_{i}^{q_i} \in \Tmodel^\ast$ for each $i \in [\numpart]$. 
Since the base model on $y_{i,j}^{\edgel,q_i}$ does not intersect with $\vermodel_{i,j}^{\edgel,q_i}, \himodel_{i}^{q_i} \in \Tmodel^\ast$, an edge model $\edgemodel_{i,j}^{s,q'}\in \Tmodel^\ast$ intersects the base model on $y_{i,j}^{\edgel,q_i}$ (assuming $i < j$), and hence $s \ge q_i$.
Since $s \le q_i$ as in the proof of \Cref{applem:heighteh}, we have $s = q_i$, that is, $\edgemodel_{i,j}^{q_i,q'}\in \Tmodel^\ast$.
Consider the path $P_{j,i}^\edgel$.
As with the above discussion, the base model on $y_{j,i}^{\edgel,q_j}$ must intersect with the edge model $\edgemodel_{i,j}^{q_i,q'}\in \Tmodel^\ast$, which implies that $q' = q_j$.
Recall that $\edgemodel_{i,j}^{q_i,q_j}\in \Tmodel$ if and only if $v_i^{q_i} v_j^{q_j} \in E_{i,j}$.
Therefore, $C$ is a clique of $G'$.
Since $|C \cap V_i| =1$ for each $i\in [\numpart]$, $(G', \numpart, (V_1, V_2, \ldots, V_\numpart))$ is a yes-instance of \textsc{Multicolored Clique}.
This completes the proof of \Cref{thm:leafage}.

\section{XP algorithms parameterized by asteroidal number} \label{sec:main}
We begin by proving \Cref{thm:fair_member} under the assumption that \Cref{thm:domset_bound,thm:dom_OPT} hold.
Given a graph $G$ and an integer $\ell$, if $\ell \ge 7\an(G) -6$, then \textsc{Minimum Membership Dominating Set} is trivially solvable by \Cref{thm:dom_OPT}.
Suppose that  $\ell \le 7\an(G) -7$. 
We apply \Cref{thm:domset_bound} to \textsc{Min.\ $(\{0,\ldots, \ell-1\}, \{1,\ldots, \ell\})$-Dominating Set}.
Since $\dmax = \ell \le 7\an(G) -7$, this yields an XP algorithm parameterized solely by the asteroidal number.
As the problem generalizes \textsc{Minimum Membership Dominating Set}, this concludes the proof of \Cref{thm:fair_member}.
In the remainder of this section, we give proofs of \Cref{thm:domset_bound,thm:dom_OPT}.

\subsection{Proof of \Cref{thm:domset_bound}} \label{subsec:domset_bound}

The case $\rho_{\max} = 0$ is trivial.
In the following, suppose that $\rho_{\max} \ge 1$.

First, arbitrarily choose a vertex $s$ from $G$ and compute $\dist{G}{s}{v}$ for every vertex $v \in V(G)$.
For a non-negative integer $i$, let $\tier_i = \{ v \in V(G) : \dist{G}{s}{v} = i \}$.
We denote by $\ecc$ the maximum distance from $s$ in $G$.
Note that $\ecc$ has a finite value as $G$ is assumed to be connected.
For an integer $i \in \{0,1, \dots, \ecc\}$, we denote by $\DPsubG{i}$ the subgraph of $G$ induced by $\tier_0 \cup \tier_1 \cup \dots \cup \tier_i$.

We will construct an XP algorithm that solves the problem parameterized by $\an(G)+\dmax$ using a dynamic programming approach.
The following lemma is essential to achieve this.

\begin{restatable}{lemma}{lemasteroidaltier} \label{lem:asteroidal_tier}
    Let $\sigma$ and $\rho$ be finite subsets of $\natnum$ with $\rho_{\max} \ge 1$, and let $\dmax = \max\{\sigma_{\max}, \rho_{\max} \}$.
    If a graph $G$ has a $(\sigma, \rho)$-dominating set $D$, then $|\tier_i \cap D| \le (\an(G) \cdot \dmax+1)(\sigma_{\max}+1)-1$ for each $i \in \{0,\dots, \ecc\}$.
\end{restatable}

\begin{proof}
    For $i=0$, clearly the lemma holds.
    For the sake of contradiction, assume that $|\tier_i \cap D| \ge (\an(G) \cdot \dmax+1)(\sigma_{\max}+1)$ for some $i\in [\ecc]$.
    We denote by $H$ the subgraph of $G$ induced by $\tier_i \cap D$.
    Since $|\neio{G}{v}\cap D| \le \sigma_{\max}$ for each $v \in D$, we have $\maxdeg{H} \le \sigma_{\max}$.
    By the well-known fact that the independence number $\alpha(H)$ of a graph $H$ satisfies $\alpha(H) \ge |V(H)|/(\maxdeg{H}+1)$, we have $\alpha(H) \ge \an(G) \cdot \dmax +1$.
    Consider an independent set $I \subseteq \tier_i \cap D$ of size $\an(G) \cdot \dmax +1$.
    Let $\DPsubG{i}^\prime$ be the graph obtained by removing all vertices in $\tier_i \setminus I$ from $\DPsubG{i}$.

    Here, we provide the following claim.

\begin{claim} \label{claim:asteroidal_largeset}
    Let $j$ be a positive integer.
    If $|I| \ge (j-1)\cdot \dmax+1$, then there is an independent set $I' \subseteq I$ of $\DPsubG{i}^\prime$ such that $|I'| = j$ and no two distinct vertices $x,y \in I'$ satisfy $\neio{\DPsubG{i}^\prime}{x} \subseteq \neio{\DPsubG{i}^\prime}{y}$ or $\neio{\DPsubG{i}^\prime}{y} \subseteq \neio{\DPsubG{i}^\prime}{x}$. Moreover, $I'$ is an asteroidal set of $\DPsubG{i}^\prime$.
\end{claim}

\begin{proof}

We prove this claim in two steps. 
\begin{enumerate}[(a)]
    \item Let $j$ be a positive integer and let $J \subseteq I$.
    If $|J| \ge (j-1)\cdot \dmax+1$, then there is an independent set $I' \subseteq J$ of $\DPsubG{i}^\prime$ such that $|I'| = j$ and no two distinct vertices $x,y \in I'$ satisfy $\neio{\DPsubG{i}^\prime}{x} \subseteq \neio{\DPsubG{i}^\prime}{y}$ or $\neio{\DPsubG{i}^\prime}{y} \subseteq \neio{\DPsubG{i}^\prime}{x}$.
    \item  If there is an independent set $I' \subseteq I$ of $\DPsubG{i}^\prime$ such that no two distinct vertices $x,y \in I'$ satisfy $\neio{\DPsubG{i}^\prime}{x} \subseteq \neio{\DPsubG{i}^\prime}{y}$ or $\neio{\DPsubG{i}^\prime}{y} \subseteq \neio{\DPsubG{i}^\prime}{x}$, then $I'$ is an asteroidal set of $\DPsubG{i}^\prime$.
\end{enumerate}

    \noindent \textbf{Claim~(a).}
        We prove the claim by induction on $j$.
        The case $j=1$ is trivial.
        Suppose that $j>1$.
        Let $z$ be a vertex in $J$ with the minimum degree in $\DPsubG{i}^\prime$ and let $u$ be a vertex in $\neio{\DPsubG{i}^\prime}{z}$, which belongs to $\tier_{i-1}$.
        Note that $\neio{\DPsubG{i}^\prime}{x} \neq \emptyset$ and $\neio{\DPsubG{i}^\prime}{x} \subseteq \tier_{i-1}$ for every $x \in I$; otherwise, we would have $\dist{G}{s}{x} \neq i$.
        We next define $\widehat{I} = \{ z' \in J : \neio{\DPsubG{i}^\prime}{z}  \subseteq \neio{\DPsubG{i}^\prime}{z'}  \}$.
        Observe that $|\widehat{I}| \le \dmax$. 
        Otherwise, the vertex $u$ would be adjacent to all vertices in $\widehat{I}$ and hence $|\neio{G}{u} \cap D| \ge |\neio{G}{u} \cap \widehat{I}| > \dmax$, contradicting that $D$ is a $(\sigma, \rho)$-dominating set.
        We have $|J \setminus \widehat{I}| \ge (j-2) \cdot \dmax+1$ and therefore, by the induction hypothesis applied to $J \setminus \widehat{I}$, $\DPsubG{i}^\prime$ has an independent set $I' \subseteq J\setminus \widehat{I}$ such that $|I'| = j-1$ and no two distinct vertices $x,y \in I'$ satisfy $\neio{\DPsubG{i}^\prime}{x} \subseteq \neio{\DPsubG{i}^\prime}{y}$ or $\neio{\DPsubG{i}^\prime}{y} \subseteq \neio{\DPsubG{i}^\prime}{x}$.
        Moreover, for each $x \in I'$, we have $\neio{\DPsubG{i}^\prime}{z} \setminus \neio{\DPsubG{i}^\prime}{x} \neq \emptyset$ due to $x \notin \widehat{I}$.
        We also have $\neio{\DPsubG{i}^\prime}{x} \setminus \neio{\DPsubG{i}^\prime}{z} \neq \emptyset$; otherwise, $x \in \widehat{I}$ or $\degv{\DPsubG{i}^\prime}{x} < \degv{\DPsubG{i}^\prime}{z}$, a contradiction.
        Therefore, $\neio{\DPsubG{i}^\prime}{z} \nsubseteq \neio{\DPsubG{i}^\prime}{x}$ and $\neio{\DPsubG{i}^\prime}{x} \nsubseteq \neio{\DPsubG{i}^\prime}{z}$ for each $x \in I'$.
        We conclude that $I'\cup \{ z \}$ is a desired independent set of $\DPsubG{i}^\prime$.
    
\vspace{\baselineskip}

        \noindent \textbf{Claim~(b).} 
        Consider three distinct vertices $x,y,z \in I'$. 
        Then, there are (possibly the same) vertices $u,v \in \tier_{i-1}$ such that $u \in \neio{\DPsubG{i}^\prime}{x} \setminus  \neio{\DPsubG{i}^\prime}{z}$ and $v \in \neio{\DPsubG{i}^\prime}{y} \setminus \neio{\DPsubG{i}^\prime}{z}$.
        Since $\tier_0 = \{s\}$ and $\neio{\DPsubG{i}^\prime}{x} \setminus \neio{\DPsubG{i}^\prime}{z} \neq \emptyset$, we have $i \ge 2$.
        The union of shortest paths in $G$ from $u$ and $v$ to $s$ contains a $u$-$v$ path $P$ in $\DPsubG{i-1}$ whose vertices other than $u$ and $v$ belong to $\tier_0 \cup \dots \cup \tier_{i-2}$.
        Since $z \in \tier_i$, no vertex of $P$ is adjacent to $z$.
        Therefore, $x$ and $y$ are contained in the same connected component of $\DPsubG{i}^\prime - \neic{\DPsubG{i}^\prime}{z}$.
        Consequently, $I'$ is an asteroidal set of $\DPsubG{i}^\prime$.
\end{proof}

    By \Cref{claim:asteroidal_largeset}, we conclude that $\DPsubG{i}^\prime$ has an asteroidal set of size $\an(G)+1$, which means that $\an(G)+1 \le \an(\DPsubG{i}^\prime)$.
    On the other hand, it is known that every asteroidal set in an induced subgraph of $G$ is also an asteroidal set in $G$~\cite{KloksKM97}, and hence we have $\an(\DPsubG{i}^\prime) \le \an(G)$.
    Therefore, we have $\an(G)+1 \le  \an(\DPsubG{i}^\prime) \le \an(G)$, a contradiction.
    % Combined with \Cref{lem:asteroidal_induced}, we have $\an(G)+1 \le  \an(\DPsubG{i}^\prime) \le \an(G)$, a contradiction.
\end{proof}

Using \Cref{lem:asteroidal_tier}, we construct a dynamic programming algorithm that solves \problem.
For algorithmic convenience, add a dummy set $\tier_{\ecc+1} = \emptyset$ and define $\DPsubG{\ecc+1} = \DPsubG{\ecc}$.
For each $i \in \{ 0, 1, \dots, \ecc\}$, our algorithm computes sets $D \subseteq V(\DPsubG{i+1})$ such that 
\begin{itemize}
    \item \rev{$|\neio{G}{v} \cap D| \in \sigma$ for every vertex $v \in V(\DPsubG{i}) \cap D$;} 
    \item  \rev{$|\neio{G}{v} \cap D| \in \rho$ for every vertex $v \in V(\DPsubG{i}) \setminus D$.}
\end{itemize}
As $G = \DPsubG{\ecc} = \DPsubG{\ecc+1}$, our algorithm finally outputs a solution for \problem.
As maintaining all possible sets $D$ could lead to an exponential-time algorithm, we only record the essential information.
In the algorithm, we guess the two sets $S_{i} = \tier_{i}\cap D$ and $S_{i+1} = \tier_{i+1}\cap D$ for each $i \in \{ 0, 1, \dots, \ecc \}$.
For each pair $(S_{i}, S_{i+1})$, we store a minimum set $D$ that satisfies the above conditions. 
By \Cref{lem:asteroidal_tier}, since the sizes of $S_{i}$ and $S_{i+1}$ are bounded by $(\an(G) \cdot \dmax+1)(\sigma_{\max}+1)-1$, it suffices to consider $O(n^{2(\an(G) \cdot \dmax+1)(\sigma_{\max}+1)-2})$ pairs.

Formally, we define $\mathcal{S}_i = \{ S \subseteq \tier_{i} : |S| \le (\an(G) \cdot \dmax+1)(\sigma_{\max}+1)-1\}$ for each $i \in \{ 0,\dots, \ecc+1 \}$.
For each triple $(i, S_i, S_{i+1})$, where $i \in \{ 0, 1, \dots, \ecc\}$, $S_{i} \in \mathcal{S}_{i}$, and $S_{i+1} \in \mathcal{S}_{i+1}$, let $\dpast{i}{S_i}{S_{i+1}}$ be the function that returns the minimum size of a set $D\subseteq V(\DPsubG{i+1})$ satisfying the following four conditions:
\begin{enumerate}[(1)]
    \item $S_{i} = \tier_{i}\cap D$;
    \item $S_{i+1} = \tier_{i+1}\cap D$;
    \item $|\neio{G}{v} \cap D| \in \sigma$ for every vertex $v \in V(\DPsubG{i}) \cap D$; and 
    \item  $|\neio{G}{v} \cap D| \in \rho$ for every vertex $v \in V(\DPsubG{i}) \setminus D$.
\end{enumerate}
If such a set does not exist, let $\dpast{i}{S_i}{S_{i+1}} = +\infty$.
% Note that for simplicity of description, the function $f$ is here described as recording the minimum size; however, it can be easily adapted to store the minimum set $D$ itself.
After obtaining $\dpast{i}{S_i}{S_{i+1}}$ for each possible triple $(i, S_i, S_{i+1})$, we conclude that $(G,k)$ is a yes-instance of \problem\ if and only if $\dpast{\ecc}{S_{\ecc}}{\emptyset} \le k$ for some $S_{\ecc} \in \mathcal{S}_{\ecc}$.
Here, note that $\tier_{\ecc} \cap D \in \mathcal{S}_{\ecc}$ for every $(\sigma, \rho)$-dominating set $D$ of $G$ by \Cref{lem:asteroidal_tier}.

We explain the computation of $\dpast{i}{S_i}{S_{i+1}}$.
We first consider the case $i=0$.
Recall that $\tier_0 = \{s\}$ and all the vertices in $S_1$ are adjacent to $s$.
We thus have the following:
\begin{align*}
    \dpast{0}{S_0}{S_{1}} = 
    \begin{cases}
        |S_1|+1 & \text{if $s \in S_0$ and $|S_1|\in \sigma$;} \\
        |S_1| & \text{if $s \notin S_0$ and $|S_1|\in \rho$; and} \\
        +\infty & \text{otherwise.}
    \end{cases}
\end{align*}

Consider the case $i>0$.
A triple $(S_{i-1}, S_i, S_{i+1})$ is \emph{valid} if the following two conditions are satisfied:
\begin{enumerate}[(1)]
\setcounter{enumi}{4}
    \item $|\neio{G}{v} \cap(S_{i-1} \cup S_i \cup S_{i+1})| \in \sigma$ for each $v \in S_i$; and 
    \item $|\neio{G}{v} \cap(S_{i-1} \cup S_i \cup S_{i+1})| \in \rho$ for each $v \in \tier_i \setminus S_i$.
\end{enumerate}
If there is a set $S_{i-1} \in \mathcal{S}_{i-1}$ such that $(S_{i-1}, S_i, S_{i+1})$ is valid, then we let
\begin{equation} \label{eq:DP}
    \dpast{i}{S_i}{S_{i+1}} = \min \{\dpast{i-1}{S_{i-1}}{S_{i}} : \text{$S_{i-1} \in \mathcal{S}_{i-1}$, $(S_{i-1}, S_i, S_{i+1})$ is valid} \} + |S_{i+1}|;
\end{equation}
otherwise, let $\dpast{i}{S_i}{S_{i+1}} = +\infty$.
% We defer the proof of correctness and analysis of running time for our algorithm to \Cref{app:DP_correctness}.

We describe the correctness of \cref{eq:DP}.
Suppose that there is a set $S_{i-1} \in \mathcal{S}_{i-1}$ such that $(S_{i-1}, S_i, S_{i+1})$ is valid.
\rev{The case $\dpast{i-1}{S_{i-1}}{S_{i}} = +\infty$ is straightforward.}
\rev{Suppose that $\dpast{i-1}{S_{i-1}}{S_{i}} < +\infty$. 
By definition, there exists a set $D'\subseteq V(\DPsubG{i})$ of size $\dpast{i-1}{S_{i-1}}{S_{i}}$ such that} 
\begin{itemize}
    \item $S_{i-1} = \tier_{i-1}\cap D'$;
    \item $S_{i} = \tier_{i}\cap D'$; 
    \item $|\neio{G}{v} \cap D'| \in \sigma$ for every vertex $v \in V(\DPsubG{i-1}) \cap D'$; and 
    \item  $|\neio{G}{v} \cap D'| \in \rho$ for every vertex $v \in V(\DPsubG{i-1}) \setminus D'$.
\end{itemize}
By Conditions~(5) and~(6), the set $D'\cup S_{i+1}$ satisfies Conditions (1)--(4) and has size $\dpast{i-1}{S_{i-1}}{S_{i}}+|S_{i+1}|$.
Taking the minimum value of $\dpast{i-1}{S_{i-1}}{S_{i}}$ among all $S_{i-1} \in \mathcal{S}_{i-1}$ for which $(S_{i-1}, S_i, S_{i+1})$ is valid, we see that $\dpast{i}{S_i}{S_{i+1}}$ is at most the right-hand side of \cref{eq:DP}.
Conversely, let $D \subseteq V(\DPsubG{i+1})$ be a set satisfying Conditions (1)--(4), and let $S_{i-1} = \tier_{i-1} \cap D$.
By Conditions (3) and (4), $D \cap V(\DPsubG{i})$ is a $(\{0,\dots,\sigma_{\max}\}, \{0,\dots,\rho_{\max}\})$-dominating set of $\DPsubG{i}$.
Moreover, $\an(\DPsubG{i}) \le \an(G)$, and the distance from $s$ to each vertex in $\DPsubG{i}$ is the same as in $G$.
Thus, \Cref{lem:asteroidal_tier} applied to $\DPsubG{i}$ implies that $S_{i-1} \in \mathcal{S}_{i-1}$.
By Conditions (3) and (4), $(S_{i-1}, S_i, S_{i+1})$ is valid.
Moreover, $D \setminus S_{i+1}$ is a candidate for $\dpast{i-1}{S_{i-1}}{S_{i}}$, and hence $|D| \ge \dpast{i-1}{S_{i-1}}{S_{i}} + |S_{i+1}|$.
This proves \cref{eq:DP}.

Suppose next that there is no set $S_{i-1} \in \mathcal{S}_{i-1}$ such that $(S_{i-1}, S_i, S_{i+1})$ is valid.
As shown above, any set $D \subseteq V(\DPsubG{i+1})$ satisfying Conditions (1)--(4) yields such a set $S_{i-1} = \tier_{i-1} \cap D$.
Thus, no such set $D$ exists, and hence $\dpast{i}{S_i}{S_{i+1}} = +\infty$.

Finally, we bound the running time of our algorithm.
For notational convenience, let $\lambda = (\an(G) \cdot \dmax+1)(\sigma_{\max}+1)-1$.
For each $i\in \{0,\dots, \ecc + 1 \}$, we have $|\mathcal{S}_i| = O(n^\lambda)$ from \Cref{lem:asteroidal_tier}.
Since $\ecc \le n$, there are $O(n^{2\lambda+1})$ possible triples $(i,S_{i},S_{i+1})$.
When $i=0$, the value of $\dpast{i}{S_i}{S_{i+1}}$ can be computed in $O(n)$ time.
When $i > 0$, the value of $\dpast{i}{S_i}{S_{i+1}}$ can be obtained in $O(n^2)$ time for each $S_{i-1} \in \mathcal{S}_{i-1}$ by counting the vertices in $\neio{G}{v} \cap(S_{i-1} \cup S_i \cup S_{i+1})$ for each $v \in \tier_i$.
Since $|\mathcal{S}_{i-1}| = O(n^\lambda)$, we conclude that the total running time of our algorithm is $O(n^{3\lambda+3}) = O(n^{3(\an(G) \cdot \dmax+1)(\sigma_{\max}+1)})$.
Note that the value $\an(G)$ used in the definition of $\mathcal{S}_i$ can be computed by the straightforward XP algorithm mentioned in the introduction.
It is clear that $\sigma_{\max} \le \dmax$, and therefore, an XP algorithm parameterized by $\an(G)+\dmax$ is obtained.
This completes the proof of \Cref{thm:domset_bound}.

\subsection{Proof of \Cref{thm:dom_OPT}} \label{subsec:dom_OPT}

% Here, we rely on the following lemma for graphs that admit a $d$-octopus. 
% (We refer the reader to \cite{FominKM04} for the definition of a $d$-octopus.)

Let $d$ be a positive integer.
We use the following lemma, which applies to graphs having a dominating set containing at most $d$ vertices in each BFS layer.

\begin{lemma}[\cite{FominKM04}] \label{lem:octopus}
    Let $G = (V, E)$ be a graph and $s$ a vertex of $G$. 
    For a non-negative integer $i$, let $\tier_i = \{ v \in V(G) : \dist{G}{s}{v} = i \}$.
    The maximum distance from $s$ is denoted by $\ecc$.
    Suppose that $G$ has a dominating set $S$ such that $|S \cap \tier_i| \le d$ for each $i \in [\ecc]$. 
    Then $G$ has a minimum dominating set $D$ such that
    \begin{equation*}
    \bigwedge_{0 \le i \le \ecc} \bigwedge_{0 \le j \le \ecc-i} \left|D \cap \bigcup_{p=i}^{i+j} \rev{\tier_p} \right| \le (j+5)d-1.
    \end{equation*}
\end{lemma}

\rev{A \emph{$d$-octopus} of $G$ is a subgraph whose vertex set forms a dominating set of $G$ and which consists of $d$ shortest paths in $G$ sharing an endpoint $s$, called its root~\cite{FominKM04}. 
Let $S$ be its vertex set. 
For the BFS layers rooted at $s$, each path contains at most one vertex in each layer, and hence $|S\cap V_i|\le d$. 
Thus, the hypothesis of \Cref{lem:octopus} is satisfied.}

\rev{If $\an(G)=1$, then $G$ is a complete graph. Choosing an arbitrary vertex $v$ gives a dominating set $D=\{v\}$ satisfying $\memD{G}{D}= 1 =7\an(G)-6$, which proves \Cref{thm:dom_OPT}.
Otherwise, $G$ has an $(\an(G)-1)$-octopus~\cite{FominKM04}. 
We thus apply \Cref{lem:octopus} with $d=\an(G)-1$ by choosing $s$ as the root of this octopus.}
Thus, $G$ has a minimum dominating set $D$ such that
\begin{equation*}
\bigwedge_{0 \le i \le \ecc} \bigwedge_{0 \le j \le \min\{2, \ecc-i\}} \left|D \cap \bigcup_{p=i}^{i+j} \rev{\tier_p} \right| \le \min\{7, \ecc-i +5 \}(\an(G)-1)-1 \le 7\an(G)-6.
\end{equation*}
Since each vertex $v$ in $\tier_i$ is adjacent only to vertices in $\tier_{i-1} \cup \tier_i \cup \tier_{i+1}$, the above formula ensures that $\memD{G}{D} \le 7\an(G)-6$.
This completes the proof of \Cref{thm:dom_OPT}.

\subsection{Applications of \Cref{thm:domset_bound}} \label{subsec:application}

We conclude this paper with a discussion of distance-$r$ variants of \textsc{Min.\ $(\sigma,\rho)$-Dominating Set}.
Let $r$ be a positive integer.
For a graph $G$, we define $\neior{G}{r}{v} = \{ u \in V(G) : 1 \le \dist{G}{u}{v} \le r \}$.
Given a graph $G$ and a non-negative integer $k$, \textsc{Min.\ Distance-$r$ $(\sigma,\rho)$-Dominating Set} determines whether $G$ has a vertex set $D$ of size at most $k$ satisfying the following two conditions:
\begin{enumerate}
    \item$|\neior{G}{r}{v} \cap D| \in \sigma$ for every vertex $v \in D$; and 
    \item $|\neior{G}{r}{v} \cap D| \in \rho$ for every vertex $v \in V(G) \setminus D$.
\end{enumerate}
\textsc{Min.\ Distance-$r$ $(\sigma,\rho)$-Dominating Set} generalizes \textsc{Min.\ $(\sigma,\rho)$-Dominating Set}.
Furthermore, \textsc{Min.\ Distance-$r$ $(\natnum,\natnump)$-Dominating Set} and \textsc{Min.\ Distance-$r$ $(\natnump,\natnump)$-Dominating Set} are also known as \textsc{Distance-$r$ Dominating Set} and \textsc{Distance-$r$ Total Dominating Set}, respectively.
% which are widely studied problems in the field of graph algorithms.

% \Cref{thm:main,thm:domset_bound} can be extended to the distance-$r$ variants as follows.
The \emph{$r$-th power} of a graph $G$, denoted by $G^r$, is a graph with the vertex set $V(G^r) = V(G)$ and edge set $E(G^r) = \{ uv  : u, v \in V(G), 1 \le \dist{G}{u}{v} \le r \}$.
Observe that \textsc{Min.\ Distance-$r$ $(\sigma,\rho)$-Dominating Set} for a graph $G$ is equivalent to \textsc{Min.\ $(\sigma,\rho)$-Dominating Set} for the graph $G^r$.
Furthermore, it is known that $\an(G^r) \le \an(G)$~\cite{HoCW00}.
Therefore, from \Cref{thm:domset_bound}, the following theorem is obtained.

\begin{theorem} \label{thm:domset_bound_r}
    Let $r$ be a fixed positive integer.
    Let $\sigma$ and $\rho$ be finite subsets of $\natnum$, and let $\dmax = \max\{\sigma_{\max}, \rho_{\max} \}$.
    Then there exists an XP algorithm for \textsc{Min.\ Distance-$r$ $(\sigma, \rho)$-Dominating Set} parameterized by asteroidal number plus $\dmax$.
\end{theorem}

\bibliography{reference}

\end{document}